\documentclass[aps,a4paper,twocolumn,superscriptaddress,longbibliography,floatfix]{revtex4-1}
\usepackage[T1]{fontenc}
\usepackage{newtxtext}
\usepackage{booktabs} 
\usepackage{amsmath}
\usepackage{mathtools}
\usepackage{amsfonts}
\usepackage{bm}
\usepackage{booktabs}
\usepackage{makecell}
\usepackage{color}
\usepackage{xcolor}
\definecolor{blue}{rgb}{0.2, 0.3, 0.85}
\definecolor{red}{rgb}{0.95, 0.1, 0.15}
\usepackage[colorlinks,bookmarks=false,citecolor=blue,linkcolor=blue,urlcolor=blue]{hyperref}

\renewcommand{\section}[1]{\noindent{\textit{\textbf{#1---}}}}
\renewcommand{\subsection}[1]{{\textit{#1.~}}}

\def\be{\begin{equation}}
\def\ee{\end{equation}}
\def\bea{\begin{eqnarray}}
\def\eea{\end{eqnarray}}

\begin{document}
\title{Topological order in random interacting Ising-Majorana chains stabilized by many-body localization}
\author{Nicolas Laflorencie}
\affiliation{Laboratoire de Physique Th\'eorique, Universit\'e de Toulouse, CNRS, UPS, France}
\affiliation{Donostia International Physics Center, Paseo Manuel de Lardizabal 4, E-20018 San Sebasti\'an, Spain}
\author{Gabriel Lemari\'e}
\affiliation{Laboratoire de Physique Th\'eorique, Universit\'e de Toulouse, CNRS, UPS, France}
\affiliation{MajuLab, CNRS-UCA-SU-NUS-NTU International Joint Research Unit, Singapore}
\affiliation{Centre for Quantum Technologies, National University of Singapore, Singapore}
\author{Nicolas Mac\'e}
\affiliation{Laboratoire de Physique Th\'eorique, Universit\'e de Toulouse, CNRS, UPS, France}
\begin{abstract}
We numerically explore $\mathbb Z_2$-symmetric random interacting Ising/Majorana chains at high energy. A very rich phase diagram emerges with two topologically distinct many-body localization (MBL) regimes separated by a much broader thermal phase than previously found. This is a striking consequence of the avalanche theory.
We further find MBL spin-glass order always associated to a many-body spectral pairing, presumably signalling a strong zero mode operator which opens fascinating perspectives for MBL-protected topological qubits.
\end{abstract}
\maketitle
\section{Introduction}
Many-body localization (MBL) in quantum interacting systems has attracted a lot of attention over the past two decades~\cite{altshuler_quasiparticle_1997,jacquod_emergence_1997,gornyi_interacting_2005,basko_metalinsulator_2006,znidaric_many-body_2008,pal_many-body_2010,bardarson_unbounded_2012,luitz_many-body_2015,alet_many-body_2018,abanin_many-body_2019}. While the first (analytical) studies addressed the fate of the non-interacting Anderson insulator against weak interactions~\cite{gornyi_interacting_2005,basko_metalinsulator_2006}, most of the subsequent numerical studies have focused on strongly interacting one-dimensional (1D) models, such as the random-field Heisenberg chain~\cite{pal_many-body_2010,luitz_many-body_2015}, for which there is a global consensus for an ergodicity-breaking transition~\cite{luitz_many-body_2015,doggen_many-body_2018,chanda_time_2020,sierant_thouless_2020,abanin_distinguishing_2021}.
In addition to convincing experimental observations~\cite{schreiber_observation_2015,smith_many-body_2016,choi_exploring_2016,roushan_spectroscopic_2017}, the existence of MBL has also been proven by Imbrie~\cite{imbrie_many-body_2016} for interacting random Ising chains governed by ${\cal{H}}_{\rm Imbrie}={\cal{H}}_{\rm TFI}+{\cal V}_{x}$. Here, 
\be
{\cal{H}}_{\rm TFI}=\sum_{i}J_i\sigma^x_i \sigma^x_{i+1}+h_i \sigma^z_i
\label{eq:TFI}
\ee
is the non-interacting transverse-field Ising (TFI) chain model, further perturbed by interactions of the form ${\cal V}_{x}=\sum_i\Gamma_i^x \sigma^x_i$ that explicitly breaks the $\mathbb{Z}_2$  symmetry of the TFI model.

So far, a few works~\cite{pekker_hilbert-glass_2014,kjall_many-body_2014,sahay_emergent_2021,moudgalya_perturbative_2020,wahl2021local} have considered MBL physics with $\mathbb{Z}_2$-preserving interactions, i.e. where the parity operator  ${{\mathbb P}}=\prod_i \sigma_i^z$ commutes with the interacting Hamiltonian. However, the parity symmetry potentially allows for  topologically distinct MBL regimes: (i) a featureless localized phase with unbroken parity 
and 
(ii) a $\mathbb{Z}_2$ broken quantum spin-glass order, for which the observation of MBL-protected topological order at all energies associated with Majorana edge states has remained elusive since the seminal work of Huse {\it et al.}~\cite{huse_localization-protected_2013}.

Very recently, two simultaneous works~\cite{sahay_emergent_2021,moudgalya_perturbative_2020} have numerically studied the high-energy properties of the ${\mathbb{Z}}_2$-symmetric random interacting Ising/Majorana (IM) chain model
\be
{\cal{H}}_{\rm IM}=\sum_{i}J_i\sigma^x_i \sigma^x_{i+1}+h_i \sigma^z_i+g\left(\sigma^z_i \sigma^z_{i+1}+\sigma^x_i \sigma^x_{i+2}\right).
\label{eq:IM}
\ee
They detected  two distinct MBL phases, but without evidence of topological order such as spectral pairing~\cite{fendley_parafermionic_2012}, separated by an intervening ergodic region which was further argued~\cite{moudgalya_perturbative_2020} to shrink into a single point of infinite randomness criticality (IRC)~\cite{fisher_critical_1995} in the limit of vanishing interaction strength.

Despite the strong interest triggered by these studies, some important issues remain widely open, raising two crucial questions: (i) what does the avalanche theory of MBL transitions~\cite{de_roeck_stability_2017} imply for this class of systems\,? and (ii) can MBL help to stabilize coherent topological qubits~\cite{bahri_localization_2015,PhysRevX.7.041062}\,?

\begin{figure}[b!]
   \centering
   \includegraphics[width=.95\columnwidth,clip]{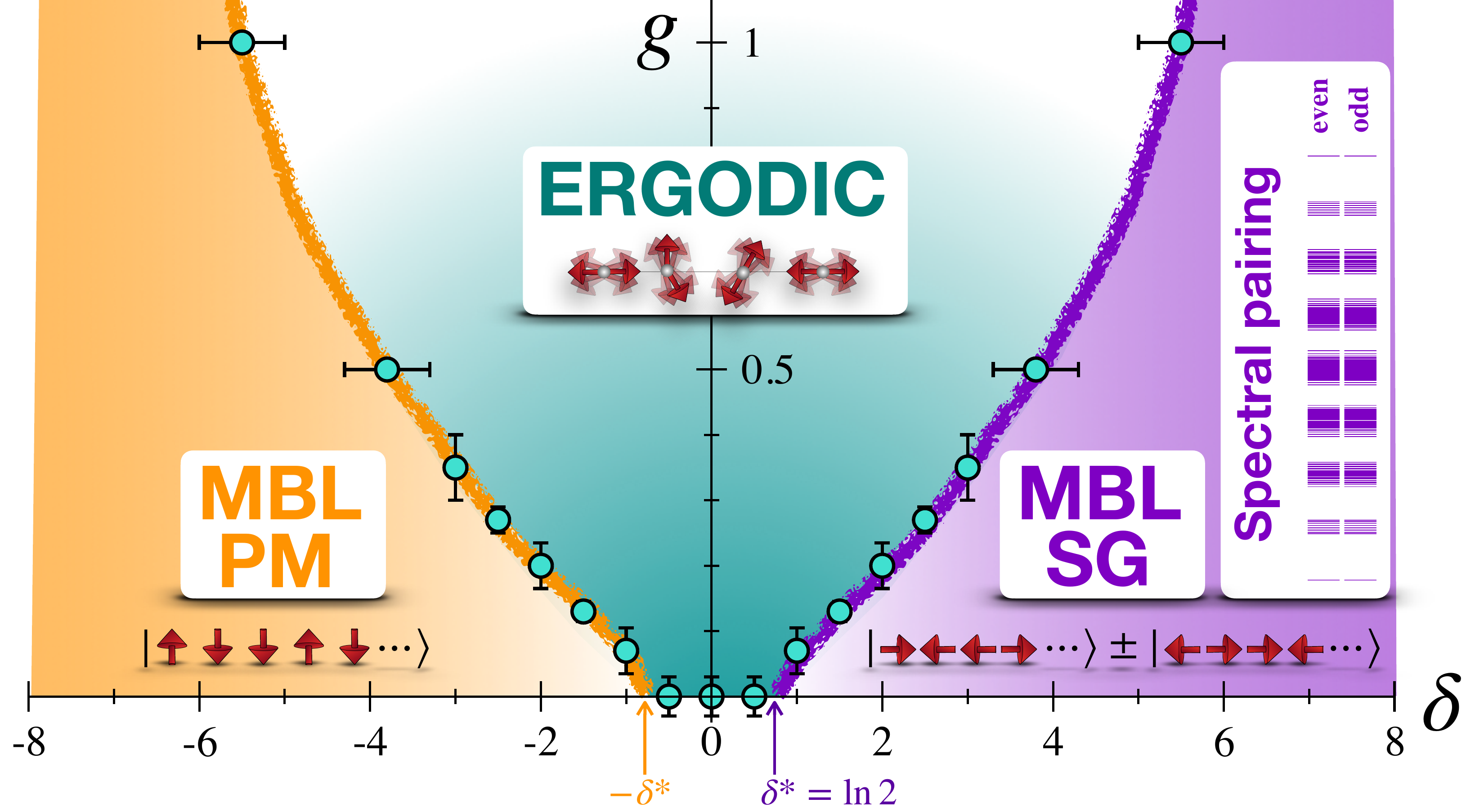}\vskip -0.35cm
      \caption{Infinite-temperature phase diagram of the random interacting Ising-Majorana chain model Eq.~\eqref{eq:IM}, obtained using shift-invert exact diagonalization.
      In the plane $\delta - g$ ($\delta={\overline{\ln J}}-{\overline{\ln \,h}}$, $g$ the interaction), a broad ergodic phase intervenes between two MBL {regimes:} a featureless paramagnet (MBL PM) 
      and a spin-glass order (MBL SG) with topological order signalled by a many-body spectral pairing. For $
      |\delta|<\delta^*=\ln 2$,  
  the non-interacting localization length exceeds the avalanche threshold  $\xi^*=1/\delta^*$ (see text). 
      }
   \label{fig:phase_diagram}
\end{figure}

\section{Main results} In order to address these critical questions, we build on shift-invert exact diagonalization to explore the ${\mathbb{Z}}_2$-symmetric disordered {\it{and}} interacting Ising-Majorana (IM) chain model Eq.~\eqref{eq:IM}
for which we provide the infinite-temperature phase diagram in Fig.~\ref{fig:phase_diagram}.  The Kramers-Wannier duality~\cite{kramers_statistics_1941} leads to symmetric phase boundaries with two different MBL regimes,
in agreement with earlier works~\cite{sahay_emergent_2021,moudgalya_perturbative_2020}. However, the global shape reported in Fig.~\ref{fig:phase_diagram} sharply contrasts with those previous findings: we observe the opening of a broad ergodic window in the $g\to 0$ limit, as a direct consequence of the avalanche criterion when applied to the non-interacting typical localization length $\xi>\xi^*\equiv(\ln 2)^{-1}$~\cite{de_roeck_stability_2017}.

We also demonstrate that the MBL quantum SG order exhibits cat-states for all energies, with a global double degeneracy of the many-body spectrum between the two parity sectors.
This is an example of MBL-protected topological order with localized Majorana edge states, and we remarkably identify a unique phase having both MBL SG {\it{and}} a paired spectrum, and hence a single transition towards ergodicity occurring when the typical localization length also exceeds the avalanche threshold $\xi^*$. This global parity degeneracy is presumably associated to a strong zero mode (SZM) operator~\cite{fendley_parafermionic_2012,noteSZM}, which opens the possibility to realize MBL-protected topological qubits with infinite coherence time at any temperature.

\vskip 0.2cm\section{Interacting Ising-Majorana chain model} Before presenting our numerical results, we first discuss some key properties of the IM Hamiltonian,  Eq.~\eqref{eq:IM}. As previously mentioned, this model preserves the $\mathbb Z_2$ parity symmetry, $[{\cal{H}}_{\rm IM},\,\mathbb P]=0$, such that one can independently diagonalize ${\cal{H}}_{\rm IM}$ in each (even and odd) parity sectors. As also exploited in Refs.~\cite{sahay_emergent_2021,moudgalya_perturbative_2020}, there is a duality transformation which ensures a symmetric phase boundary under $\delta={\overline{\ln\left({J}/{h}\right)}}\to -\delta$~\cite{kramers_statistics_1941}, as seen in Fig.~\ref{fig:phase_diagram}.

If one rewrites Pauli operators as Majorana fermions: $\gamma_{2j-1}=\sigma_j^x\prod_{\ell< j}\sigma_\ell^z$, and $\gamma_{2j}=\sigma_j^y\prod_{\ell<j}\sigma_\ell^z$, the original IM spin model reads 
\be
{\cal{H}}_{\rm IM}=\sum_k-{\rm{i}}t_k\gamma_k\gamma_{k+1}-g\gamma_k\gamma_{k+1}\gamma_{k+2}\gamma_{k+3},
\ee
with hopping terms $t_{2j-1}=h_j$, $t_{2j}=J_{j}$, and interaction strength $g$. Ground-state properties have been obtained in some limiting cases, such as the random non-interacting chain by Fisher~\cite{fisher_critical_1995}, and for the interacting problem in the absence~\cite{rahmani_phase_2015} or presence of disorder~\cite{lobos_interplay_2012,crepin_nonperturbative_2014,gergs_topological_2016,karcher_disorder_2019}. Interestingly, the Majorana edge states localized at the boundaries of a non-interacting open chain~\cite{kitaev_unpaired_2001,fendley_parafermionic_2012}  in the topological regime ($\delta>0$) appear robust against disorder or weak interactions~\cite{gergs_topological_2016}. On the other hand, IRC at $\delta=0$ was found~\cite{karcher_disorder_2019} unstable towards localization and topological ordering for $g>0$ (repulsive interactions), while attraction is also relevant but could apparently drive IRC to a different (Ising) criticality~\cite{karcher_disorder_2019}.

Anyhow, at high energy the interaction sign is expected to be irrelevant
\cite{lin_many-body_2018}, such that we work with $g>0$, assuming similar results for attractive couplings. We then take box distributions for both  couplings $J_i$ and fields $h_i$, $P_{J/h}={\rm{Box}}[0\,,W_{J/h}]$ being uniform between $0$ and $W_{J/h}$, with $W_J=W_{h}^{-1}=W$ so that the control parameter is $\delta={\overline{\ln J}}-{\overline{\ln \,h}}=2\ln W$. In the rest of the paper, we present and discuss in details the results of numerical simulations that led us to the  phase diagram shown in Fig.~\ref{fig:phase_diagram}. Using the shift-invert ED technique~\cite{luitz_many-body_2015,pietracaprina_shift-invert_2018}, we have studied open IM chains Eq.~\eqref{eq:IM} up to $L=16$ sites~\footnote{This corresponds to 32 Majorana fermions, and Hilbert spaces of maximum size 65\,536.}, and computed $\sim 100$ mid-spectrum (infinite-temperature) eigenpairs for each parity sector ($p=\pm 1$), for $\sim 500-1000$ independent random samples. 
\vskip 0.2cm
\section{Emergent ergodicty between distinct MBL phases} 
\begin{figure}[t!]
   \centering
   \includegraphics[width=\columnwidth,clip]{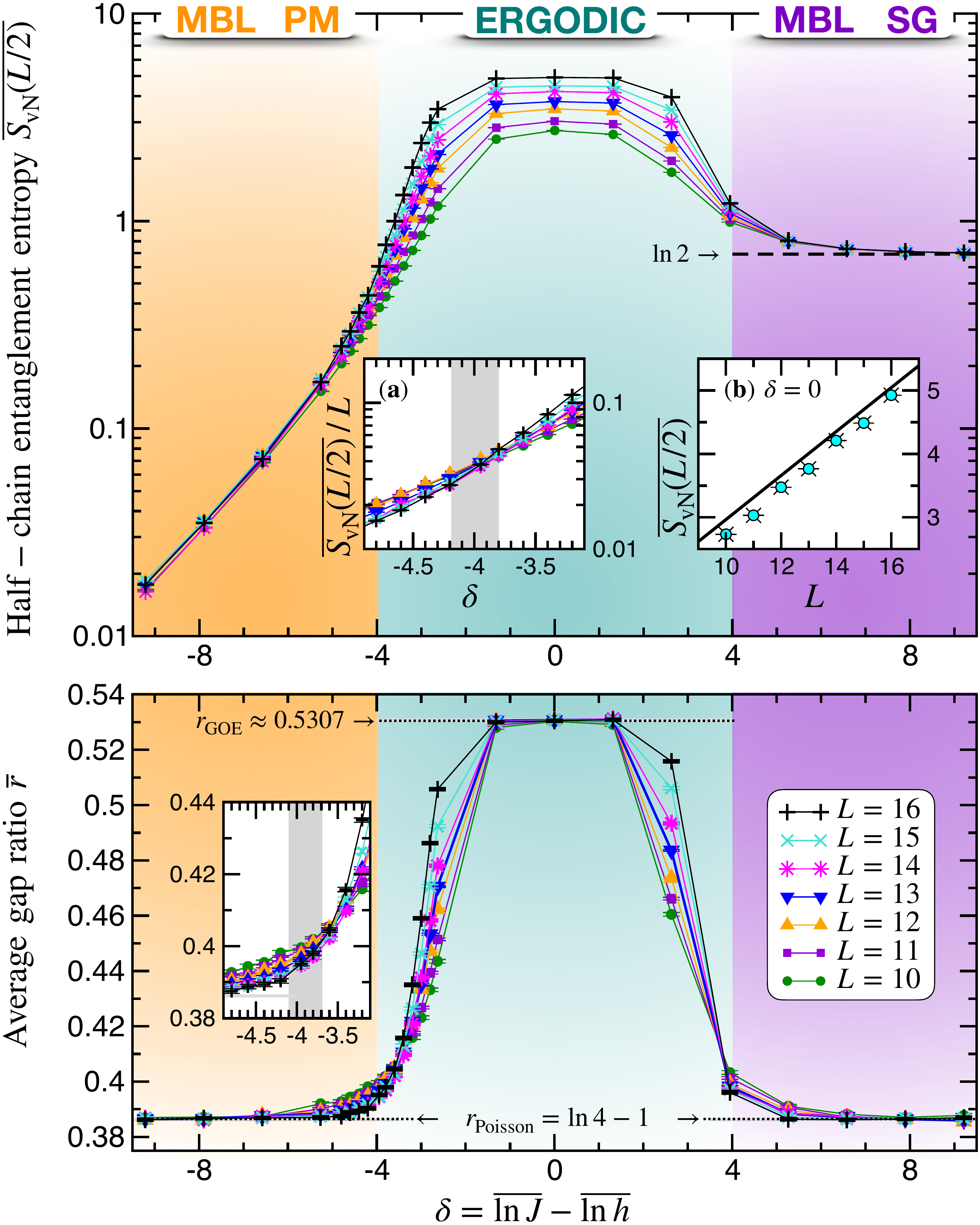}
      \caption{Infinite-temperature ED results for the IM model Eq.~\eqref{eq:IM} at interaction strength $g=0.5$, showing the sequence MBL PM ---  Ergodic --- MBL SG upon varying $\delta$ (horizontal scan in Fig.~\ref{fig:phase_diagram}). Top panel: the half-chain von-Neumann entropy allows for a clear qualitative distinction between area and volume-law entanglement (Inset (b) shows $\delta=0$ data, and the line is the infinite-temperature expectation $\ln 2 \times  L/2-1/2$), with a transition signalled by a crossing for $S_{\rm vN}/L$, see inset (a). Note also the asymmetry upon the duality mapping $\delta\to -\delta$ (see text). 
      Bottom panel: the gap ratio ${\overline{r}}$ which respects the dual symmetry, clearly displays Poisson {\it{vs.}} GOE statistics. Inset: zoom in the transition region at $\delta\approx -4$, in perfect agreement with $S_{\rm vN}$ data, see inset top (a).}
   \label{fig:g05}
\end{figure}

We first illustrate in  Fig.~\ref{fig:g05} the sequence of three phases which appear for finite interaction: MBL PM --- Ergodic --- MBL SG, upon varying $\delta$ (corresponding to an horizontal scan in the phase diagram at $g=0.5$). Fig.~\ref{fig:g05} shows two classic estimates for these phases and associated transitions: the average half-chain von-Neuman entropy ${\overline{S_{\rm vN}(L/2)}}$~\footnote{For odd system sizes $L=2p+1$ we cut at the bond $x=p$.}, probing area {\it{vs.}} volume-law entanglement, and the average (parity-resolved)  gap-ratio ${\overline{r}}$~\cite{oganesyan_localization_2007,atas_distribution_2013} which diagnoses a global change in the spectral statistics: Poisson {\it{vs.}} GOE, occurring in each parity sector. A broad thermal (ergodic) phase emerges for $\left|\delta\right|\lesssim 4$ in between two regimes showing MBL physics with uncorrelated Poisson spectral statistics and area-law entanglement. Note however that in contrast with the gap ratio, the entanglement does not respect dual symmetry $\delta \to -\delta$. Indeed, $S_{\rm vN}$ vanishes at large negative $\delta$, while $S_{\rm vN}\to \ln 2$ at large positive $\delta$. This signals a quantum-order with a "cat-state" structure for the eigenstates, of the form (in the $\sigma^x$ basis) 
$|\phi\rangle\sim |\hskip -0.05cm \uparrow\downarrow\downarrow\cdots \uparrow\uparrow\rangle_{x}+p|\hskip -0.05cm \downarrow\uparrow\uparrow\cdots \downarrow\downarrow\rangle_{x}$, where $p=\pm 1$ is the parity sector. In addition, we expect such even and odd eigenstates to be pair-wise degenerate across the full many-body spectrum, as we will elaborate on further below. 
\

\begin{figure*}[!t]
\centering
\includegraphics[width=1.485\columnwidth,clip]{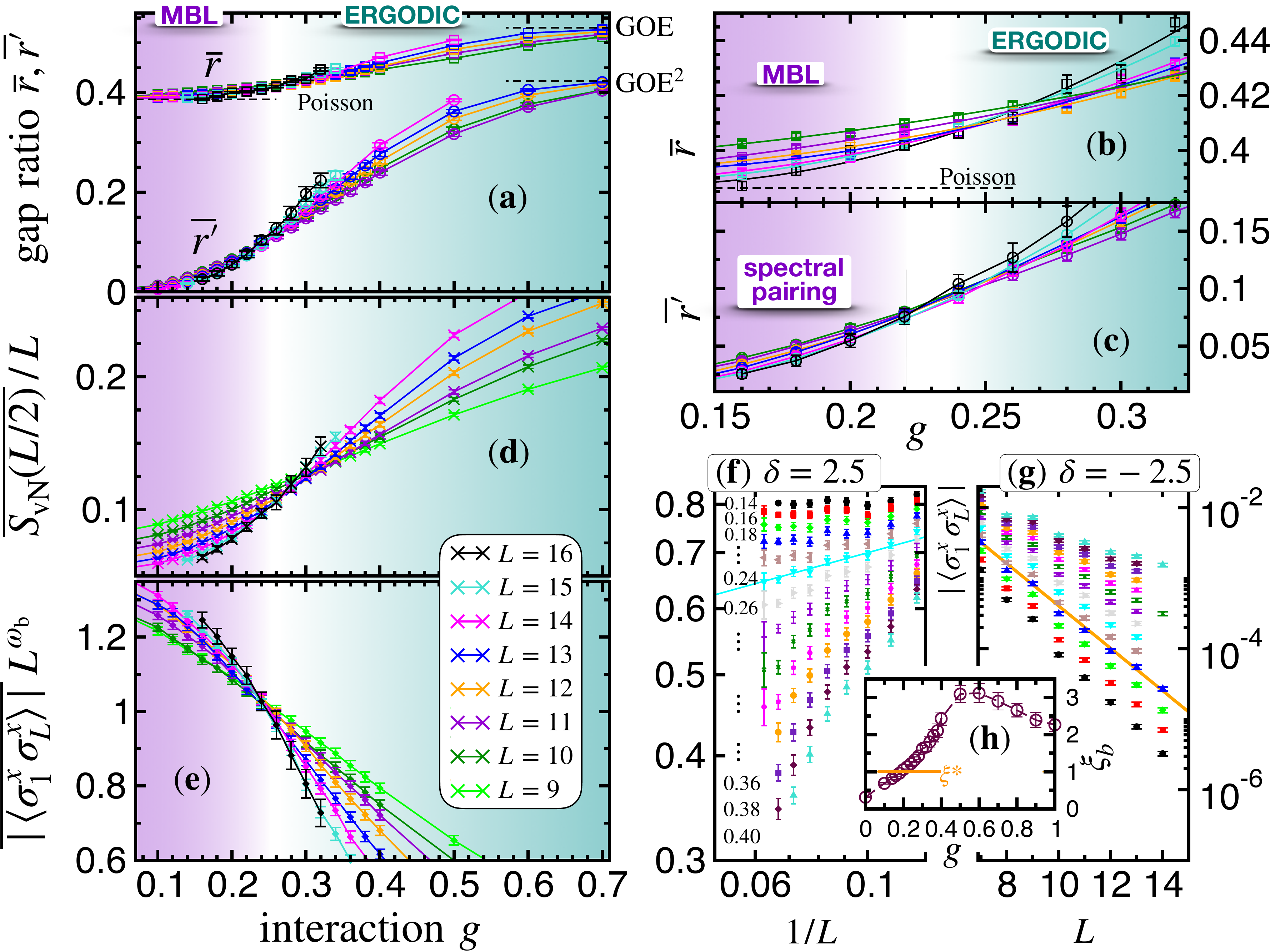}
\caption{Infinite-temperature ED results for the IM model Eq.~\eqref{eq:IM} at $\delta=2.5$ (vertical scan in  Fig.~\ref{fig:phase_diagram}, except for panel (g) at $\delta=-2.5$). All estimates (a-e) agree for an MBL SG --- thermal transition for $g_c\sim 0.23$ (see text). Panel (f): log-log plot of the average boundary correlators ${\overline{|\langle \sigma_1^x\,\sigma_L^x\rangle|}}$ for various values of $g$ (as indicated on the plot), with a critical decay best described by a power-law with an exponent  $\omega_b=0.17$ (line). Panels (b) and (c) show a zoom of panel (a) for the two gap ratios in the critical region. Panel (h): numerical estimate for the localization length $\xi_{\rm b}$ (in units of $\ln 2$) extracted from the end-to-end correlations at $\delta=-2.5$, displayed in panel (g) for the same values of the interaction strength as shown in panel (f). The avalanche condition is found $\xi_{\rm b}=\xi^*=(\ln 2)^{-1}$ (orange line) at criticality ($g\sim 0.2$), while deep in the thermal regime we observe the ETH result $\xi_{\rm b}\to 2\xi^*$, expected at large $g$.}
   \label{fig:delta2.5}
\end{figure*}

\vskip 0.2cm 
\section{Topological order, strong zero mode, and spectral pairing} For finite size systems, spontaneous breaking of the $\mathbb Z_2$ symmetry (associated to magnetic order with $\langle\sigma_i^x\rangle\neq 0$ in the thermodynamic limit) is usually detected using spin-spin correlation functions, as discussed in the supplemental material~\cite{sm}. Remarkably, this local magnetic order takes a non-local form in terms of fermions, as first observed by Kitaev~\cite{kitaev_unpaired_2001} for clean non-interacting chains. Indeed, in the ordered phase ($J>h$), free fermions display non-trivial topological properties, with "unpaired" Majorana edge states: the so-called SZM~\cite{fendley_parafermionic_2012} localized at the boundaries of an open chain.
For random TFI chains, when $\delta>0$, SZM operators can also be explicitly defined~\cite{fendley_parafermionic_2012,laflo_2022}, commuting with ${\cal{H}}_{\rm TFI}$ and mapping the even sector to the odd. This implies a spectral pairing at {\it{all}} anergies and infinite coherence time for the edge spins.

In contrast, the situation is more delicate in the presence of  interactions where no explicit construction of SZM is known, except for the integrable XYZ chain~\cite{Fendley_2016}. For the non-integrable disorder-free ${\cal{H}}_{\rm IM}$ Eq.~\eqref{eq:IM}, Kemp {\it al.}~\cite{Kemp_2017} have shown the existence of an "almost" SZM that almost commutes with the Hamiltonian, reducing the edge coherence to finite time. Here for the random IM chain model, the topological order appears explicitly in our numerical results through an even-odd degeneracy of the many-body spectrum (see Fig.~\ref{fig:phase_diagram}), with an exponentially small finite-size level splitting $\Delta_{\rm parity}\sim {\rm e}^{-L/\xi}$, where $\xi$ is viewed as the edge mode localization length. From a global spectroscopic point of view, this parity degeneracy competes with the many-body level spacing $\Delta_{\rm mb}\sim {\rm{e}}^{-sL}$ ($s$ is the entropy density) such that spectral pairing is resolved only if $\Delta_{\rm parity}\ll\Delta_{\rm mb}$, {\it{i.e.}} $\xi<1/s$. 

In practice, the detection of the topological pairing is achieved using the (parity-mixed) gap ratio~\cite{huse_localization-protected_2013,vasseur_particle-hole_2016} defined by $r'_i={\rm{min}}({\Delta_i,\Delta_{i+1}})/{\rm{max}}({\Delta_i,\Delta_{i+1}})$, where the individual gaps $\Delta_i$ are computed within the many-body spectrum when both parity sectors are mixed. In the MBL topological regime, we expect ${\overline{r'}}\sim \Delta_{\rm parity}/\Delta_{\rm mb} \to 0$ if $\xi<{1}/{s}$, while  Poisson statistics should arise when $\xi>{1}/{s}$ (as well as for a non-topological MBL phase), where ${\overline{r'}}=\ln 4 -1$ takes the same value as the parity-resolved ratio ${\overline{r}}$. Interestingly, the ergodic regime also manifests in the mixed spectral statistics, where two GOE blocks yield ${\overline{r'}}{_{{\rm GOE}^2}}\approx 0.4234$, as shown recently~\cite{giraud_probing_2020}. Table~\ref{tab:1} summarizes these spectral features. 
\begin{table}[h!]
\begin{tabular}{l|c|c|c}
&Ergodic& MBL &MBL + paired spectrum\\
\hline
${\overline{r'}}_{\rm mixed}$& $0.4234$~(GOE$^2$) & $\ln 4 -1$ &${\rm{e}}^{L(s-\xi^{-1})} \to 0$\\ 
\hline
${\overline{r}}_{\rm resolved}$& $0.5307$~(GOE) & \multicolumn{2}{c}{$\ln 4 -1$~(Poisson)}\\
\bottomrule
\end{tabular}
\vskip -0.25cm 
\caption{\label{tab:1} Parity-mixed ${\overline{r'}}$ and parity-resolved ${\overline{r}}$ values for the gap ratios across the different regimes.}
\end{table}
%


Ref.~\cite{huse_localization-protected_2013} conjectured that two types of MBL orders may emerge, with and without spectral pairing. This issue is addressed 
in Fig.~\ref{fig:delta2.5} where we show $T=\infty$ ($s=\ln 2$) results for a vertical scan in the phase diagram, collected at $\delta=2.5$. Several estimates for the MBL SG --- ergodic transition are shown: (a-c) parity-resolved and parity-mixed gap ratios ;  (d) the von-Neuman entropy density ; (e-g)  end-to-end correlators. As previously observed (Fig.~\ref{fig:g05}), here also the area to volume-law entanglement transition coincides with the Poisson --- GOE change in the parity-resolved spectral statistics, both observed for $g_c=0.23\pm 0.03$. In addition, the end-to-end correlations  
$\langle \sigma_1^x\sigma_{L}^x\rangle$, expected to be a good proxy for detecting Majorana edge modes~\cite{laflo_2022}, also display a clear ordering transition   at $g_c\approx 0.24$~\footnote{We observe a critical power-law decay ${\overline{|\langle \sigma_1^x\,\sigma_L^x\rangle|}}\sim L^{-\omega_{\rm b}}$ at $g_c= 0.24$ with $\omega_{\rm b}\approx 0.17$ which contrasts with free-fermions where $\omega_b=1$~\cite{fisher_distributions_1998}.}. This presumably rules out the possibility of an intermediate MBL PM regime~\cite{huse_localization-protected_2013,kjall_many-body_2014}, a result also strongly supported by the parity-mixed gap ratio ${\overline{r'}}$ shown in Fig.~\ref{fig:delta2.5} (a,\,c). Indeed, one sees a unique MBL regime associated to spectral pairing for $g\le g_c$, followed by an ergodic phase where ${\overline{r}}$ takes its GOE value $\approx 0.5307$ together with its parity-mixed counterpart ${\overline{r'}}$ which saturates to its GOE$^2$ value $\approx 0.4234$, as expected for two GOE blocks~\cite{giraud_probing_2020}. 

The infinite-temperature pairing transition signals that the typical localization length $\xi$ (controlling the parity gap decay) has reached $\xi^*=(\ln 2)^{-1}$, a value which strikingly coincides with the  avalanche threshold~\cite{de_roeck_stability_2017}. 
At this stage, it becomes very instructive to make a small detour to the other side of the phase diagram ($\delta<0$) where there is no topological MBL order. Instead, both MBL PM and ergodic regimes have exponential decaying end-to-end correlations ${\overline{|\langle \sigma_1^x\,\sigma_L^x\rangle|}}\sim {\rm{e}}^{-L/\xi_{\rm b}}$, controlled by the localization length $\xi_{\rm b}$, as we confirm in Fig.~\ref{fig:delta2.5} (g)~\cite{notexi}. The $g$-dependence of $\xi_{\rm b}$ at $\delta=-2.5$, Fig.~\ref{fig:delta2.5} (h), remarkably establishes that the MBL PM --- ergodic transition (at $g\sim 0.2$) is again characterized by $\xi_{\rm b}=\xi^*=(\ln 2)^{-1}$, further meeting the avalanche criterion~\cite{de_roeck_stability_2017}.

\vskip 0.2cm
\section{Avalanche, infinite randomness and weak interactions} The main idea behind the avalanche instability~\cite{de_roeck_stability_2017} is that a small (rare) ergodic region embedded inside an otherwise localized system can nucleate a growing ergodic surrounding (ultimately thermalizing the whole system), only if the typical localization length $\xi>\xi^*=(\ln 2)^{-1}$,  a critical threshold fixed by the many-body spacing in the middle of the spectrum. So far, our numerics have adhered this avalanche condition, for both sides of the phase diagram, leading to the remarkable consequence that MBL SG is always accompanied by spectral pairing. 

\begin{figure}[h!]
   \centering
   \includegraphics[width=1\columnwidth,clip]{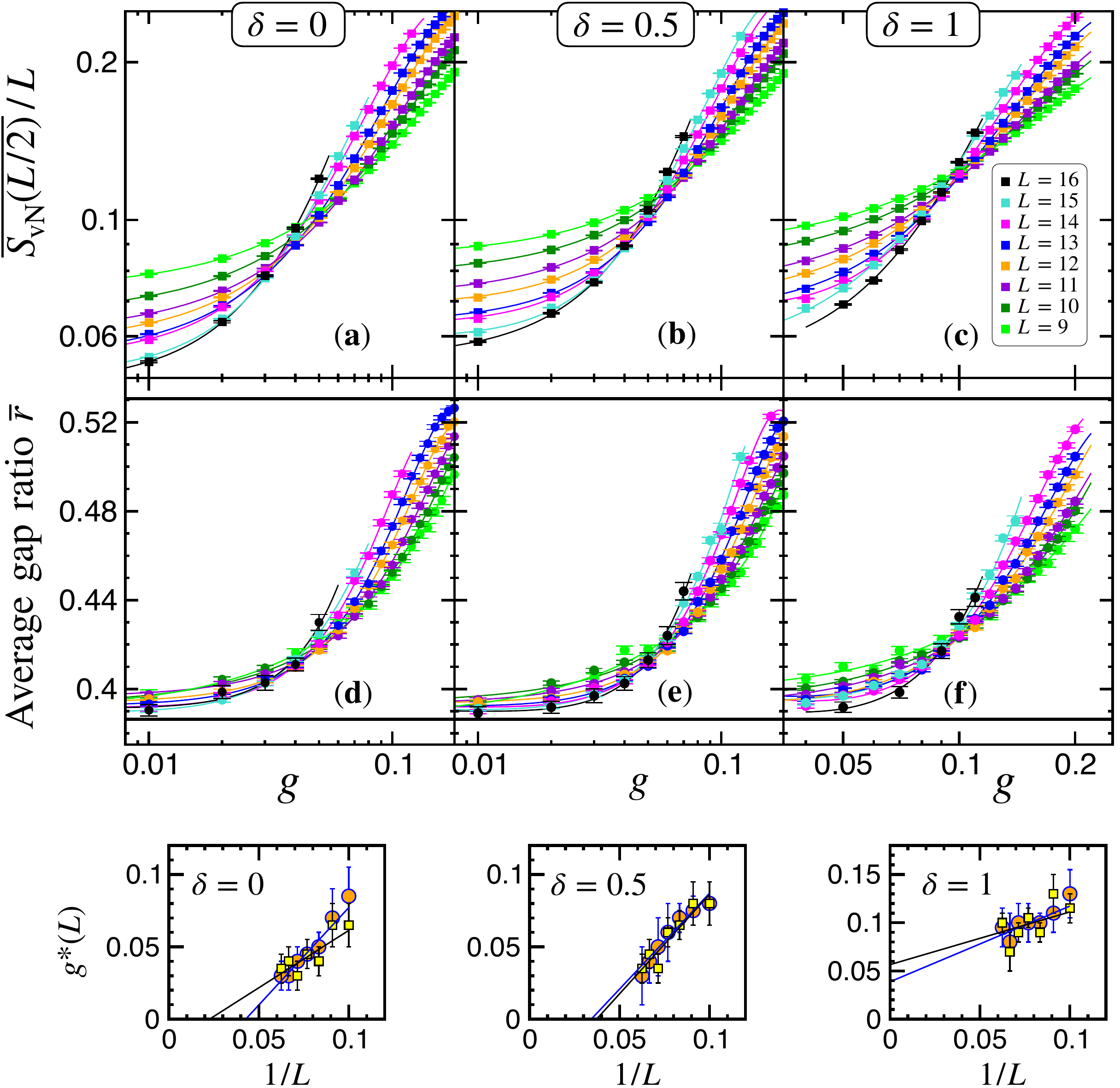}
  \caption{ED results for the IM model Eq.~\eqref{eq:IM} at $\delta=0,\,0.5,\,1$: vertical scans in  Fig.~\ref{fig:phase_diagram} for weak interactions.  Finite-size crossings of ${\overline{S_{\rm vN}(L/2)}}/L$ (panels a, b, c) and ${\overline{r}}$ (d, e, f) display visible finite-size drifts. The size-dependent drift of the crossing points $g^*(L)$ (bottom insets: squares for ${\overline{S_{\rm vN}(L/2)}}/L$ and circles for ${\overline{r}}$), supports $g_c=0$ for $\delta=0,\,0.5$ and $g_c\approx 0.05$ for $\delta=1$.}
   \label{fig:crossings}
\end{figure}

Building further on this, the condition $\xi>\xi^*$ is also satisfied without interaction~\footnote{Where the typical localization length is $1/\delta$~\cite{fisher_critical_1995}.} when $|\delta|<\delta^*\equiv \ln 2$ (See Fig.~\ref{fig:phase_diagram}).
We therefore expect ergodic instability upon infinitesimal interaction for such a region $[-\ln 2,\,\ln 2]$ surrounding IRC, thus extending the already discussed $\delta=0$ case~\cite{moudgalya_perturbative_2020}. This is checked numerically in Fig.~\ref{fig:crossings} at $\delta=0,\,0.5,\,1$, where both  ${\overline{S_{\rm vN}(L/2)}}/L$ and ${\overline{r}}$ experience strong drifts which only stop for a non-zero interaction  $g_c\approx 0.05$ for $\delta=1$. Our results for $\delta=0$ in Fig.~\ref{fig:crossings} (a,\,d) appear very similar to $\delta=0.5$, see Fig.~\ref{fig:crossings} (b,\,e), thus supporting such a scenario. As a consequence, a broad ergodic window opens in the vanishing interaction limit, thus reshaping the phase diagram of the $\mathbb Z_2$-symmetric disordered and interacting Ising-Majorana chain model Eq.~\eqref{eq:IM} shown in  Fig.~\ref{fig:phase_diagram}, contrasting with previous proposals~\cite{Parameswaran_2018,sahay_emergent_2021,moudgalya_perturbative_2020}

\vskip 0.2cm
\section{Discussions and conclusions} Our numerical investigation of the disordered interacting Ising/Majorana chain model brings the three following key results. (i) Many-body and Anderson localizations become unstable to interactions whenever the typical localization length exceeds a threshold: $\xi>\xi^*=(\ln 2)^{-1}$, thus meeting the so-called avalanche criterion.
(ii) Consequently, weak interactions destabilize not only the infinite randomness critical point towards thermalization, but also its  vicinity, as long as the typical localization length remains larger than the avalanche threshold, which is expected for $|\delta|<\delta^*\equiv \ln 2$.
(iii) Localization-protected topological order survives weak interactions and the MBL-SG regime is {\it{always}} associated to a many-body spectral pairing. 

This spectral pairing associated with MBL  has very strong implications. First of all, our results give a clear answer to a question raised in the seminal paper of Huse {\it{et al.}}~\cite{huse_localization-protected_2013}: here we unambiguously find that MBL SG order cannot occur without spectral pairing. This global parity degeneracy is presumably a smoking gun for the presence of a SZM in the disordered interacting Majorana chain model, which is a non-trivial result. Indeed, as shown for the disorder-free case~\cite{Kemp_2017}, finite interactions destroy the exactness of the zero-mode which then becomes an "almost" SZM with an enhanced but finite lifetime for edge spins, with a breakdown of spectral pairing. In our case, the robustness of the spectral pairing is a very strong indication that  disorder stabilizes the SZM. A direct consequence is that edge spins must display an infinite coherence time at any temperature, thus avoiding "quasiparticle poisoning"~\cite{PhysRevB.85.174533} thanks to MBL which may therefore help to stabilize topological qubits at finite temperature.

Nevertheless, it is important to take a critical step back on some of these conclusions, simply because our results come from numerical simulations which are far from the thermodynamic limit. This prompts us to remain careful, in particular we cannot entirely exclude the existence of an intermediate phase  in Fig.~\ref{fig:phase_diagram} where MBL SG would occur without spectral pairing, while our finite-size numerics appear more consistent with a single spectral-paired MBL SG regime. We finally note that finite-size effects associated to the avalanche instability~\cite{morningstar2021avalanches,sels2021markovian} may point towards an even wider opening of the ergodic phase at weak interaction.

\vskip 0.2cm\section{Acknowledgments}
It is a pleasure to thank D. Aceituno, J. Bardarson, L. Herviou, D. Huse, J. Kemp, and S. Mirlin for useful discussions or comments. 
This work benefited from the support of the projects GLADYS ANR-19- CE30-0013 of the French National Research Agency (ANR), the EUR grant NanoX No. ANR-17-EURE-0009 in the framework of the "Programme des Investissements d'Avenir", and by the Singapore Ministry of Education Academic Research Fund Tier I (WBS No. R-144- 000-437-114). We acknowledge the use of HPC resources from CALMIP (grants 2020-P0677 and 2021-P0677).


\begin{thebibliography}{62}%
\makeatletter
\providecommand \@ifxundefined [1]{%
 \@ifx{#1\undefined}
}%
\providecommand \@ifnum [1]{%
 \ifnum #1\expandafter \@firstoftwo
 \else \expandafter \@secondoftwo
 \fi
}%
\providecommand \@ifx [1]{%
 \ifx #1\expandafter \@firstoftwo
 \else \expandafter \@secondoftwo
 \fi
}%
\providecommand \natexlab [1]{#1}%
\providecommand \enquote  [1]{``#1''}%
\providecommand \bibnamefont  [1]{#1}%
\providecommand \bibfnamefont [1]{#1}%
\providecommand \citenamefont [1]{#1}%
\providecommand \href@noop [0]{\@secondoftwo}%
\providecommand \href [0]{\begingroup \@sanitize@url \@href}%
\providecommand \@href[1]{\@@startlink{#1}\@@href}%
\providecommand \@@href[1]{\endgroup#1\@@endlink}%
\providecommand \@sanitize@url [0]{\catcode `\\12\catcode `\$12\catcode
  `\&12\catcode `\#12\catcode `\^12\catcode `\_12\catcode `\%12\relax}%
\providecommand \@@startlink[1]{}%
\providecommand \@@endlink[0]{}%
\providecommand \url  [0]{\begingroup\@sanitize@url \@url }%
\providecommand \@url [1]{\endgroup\@href {#1}{\urlprefix }}%
\providecommand \urlprefix  [0]{URL }%
\providecommand \Eprint [0]{\href }%
\providecommand \doibase [0]{http://dx.doi.org/}%
\providecommand \selectlanguage [0]{\@gobble}%
\providecommand \bibinfo  [0]{\@secondoftwo}%
\providecommand \bibfield  [0]{\@secondoftwo}%
\providecommand \translation [1]{[#1]}%
\providecommand \BibitemOpen [0]{}%
\providecommand \bibitemStop [0]{}%
\providecommand \bibitemNoStop [0]{.\EOS\space}%
\providecommand \EOS [0]{\spacefactor3000\relax}%
\providecommand \BibitemShut  [1]{\csname bibitem#1\endcsname}%
\let\auto@bib@innerbib\@empty
\bibitem [{\citenamefont {Altshuler}\ \emph {et~al.}(1997)\citenamefont
  {Altshuler}, \citenamefont {Gefen}, \citenamefont {Kamenev},\ and\
  \citenamefont {Levitov}}]{altshuler_quasiparticle_1997}%
  \BibitemOpen
  \bibfield  {author} {\bibinfo {author} {\bibfnamefont {B.~L.}\ \bibnamefont
  {Altshuler}}, \bibinfo {author} {\bibfnamefont {Y.}~\bibnamefont {Gefen}},
  \bibinfo {author} {\bibfnamefont {A.}~\bibnamefont {Kamenev}}, \ and\
  \bibinfo {author} {\bibfnamefont {L.~S.}\ \bibnamefont {Levitov}},\ }\href
  {\doibase 10.1103/PhysRevLett.78.2803} {\bibfield  {journal} {\bibinfo
  {journal} {Phys. Rev. Lett.}\ }\textbf {\bibinfo {volume} {78}},\ \bibinfo
  {pages} {2803} (\bibinfo {year} {1997})}\BibitemShut {NoStop}%
\bibitem [{\citenamefont {Jacquod}\ and\ \citenamefont
  {Shepelyansky}(1997)}]{jacquod_emergence_1997}%
  \BibitemOpen
  \bibfield  {author} {\bibinfo {author} {\bibfnamefont {P.}~\bibnamefont
  {Jacquod}}\ and\ \bibinfo {author} {\bibfnamefont {D.~L.}\ \bibnamefont
  {Shepelyansky}},\ }\href {\doibase 10.1103/PhysRevLett.79.1837} {\bibfield
  {journal} {\bibinfo  {journal} {Phys. Rev. Lett.}\ }\textbf {\bibinfo
  {volume} {79}},\ \bibinfo {pages} {1837} (\bibinfo {year}
  {1997})}\BibitemShut {NoStop}%
\bibitem [{\citenamefont {Gornyi}\ \emph {et~al.}(2005)\citenamefont {Gornyi},
  \citenamefont {Mirlin},\ and\ \citenamefont
  {Polyakov}}]{gornyi_interacting_2005}%
  \BibitemOpen
  \bibfield  {author} {\bibinfo {author} {\bibfnamefont {I.~V.}\ \bibnamefont
  {Gornyi}}, \bibinfo {author} {\bibfnamefont {A.~D.}\ \bibnamefont {Mirlin}},
  \ and\ \bibinfo {author} {\bibfnamefont {D.~G.}\ \bibnamefont {Polyakov}},\
  }\href {\doibase 10.1103/PhysRevLett.95.206603} {\bibfield  {journal}
  {\bibinfo  {journal} {Phys. Rev. Lett.}\ }\textbf {\bibinfo {volume} {95}},\
  \bibinfo {pages} {206603} (\bibinfo {year} {2005})}\BibitemShut {NoStop}%
\bibitem [{\citenamefont {Basko}\ \emph {et~al.}(2006)\citenamefont {Basko},
  \citenamefont {Aleiner},\ and\ \citenamefont
  {Altshuler}}]{basko_metalinsulator_2006}%
  \BibitemOpen
  \bibfield  {author} {\bibinfo {author} {\bibfnamefont {D.~M.}\ \bibnamefont
  {Basko}}, \bibinfo {author} {\bibfnamefont {I.~L.}\ \bibnamefont {Aleiner}},
  \ and\ \bibinfo {author} {\bibfnamefont {B.~L.}\ \bibnamefont {Altshuler}},\
  }\href {\doibase 10.1016/j.aop.2005.11.014} {\bibfield  {journal} {\bibinfo
  {journal} {Annals of Physics}\ }\textbf {\bibinfo {volume} {321}},\ \bibinfo
  {pages} {1126} (\bibinfo {year} {2006})}\BibitemShut {NoStop}%
\bibitem [{\citenamefont {{\v Z}nidari{\v c}}\ \emph
  {et~al.}(2008)\citenamefont {{\v Z}nidari{\v c}}, \citenamefont {Prosen},\
  and\ \citenamefont {Prelov{\v s}ek}}]{znidaric_many-body_2008}%
  \BibitemOpen
  \bibfield  {author} {\bibinfo {author} {\bibfnamefont {M.}~\bibnamefont {{\v
  Z}nidari{\v c}}}, \bibinfo {author} {\bibfnamefont {T.}~\bibnamefont
  {Prosen}}, \ and\ \bibinfo {author} {\bibfnamefont {P.}~\bibnamefont
  {Prelov{\v s}ek}},\ }\href {\doibase 10.1103/PhysRevB.77.064426} {\bibfield
  {journal} {\bibinfo  {journal} {Phys. Rev. B}\ }\textbf {\bibinfo {volume}
  {77}},\ \bibinfo {pages} {064426} (\bibinfo {year} {2008})}\BibitemShut
  {NoStop}%
\bibitem [{\citenamefont {Pal}\ and\ \citenamefont
  {Huse}(2010)}]{pal_many-body_2010}%
  \BibitemOpen
  \bibfield  {author} {\bibinfo {author} {\bibfnamefont {A.}~\bibnamefont
  {Pal}}\ and\ \bibinfo {author} {\bibfnamefont {D.~A.}\ \bibnamefont {Huse}},\
  }\href {\doibase 10.1103/PhysRevB.82.174411} {\bibfield  {journal} {\bibinfo
  {journal} {Phys. Rev. B}\ }\textbf {\bibinfo {volume} {82}},\ \bibinfo
  {pages} {174411} (\bibinfo {year} {2010})}\BibitemShut {NoStop}%
\bibitem [{\citenamefont {Bardarson}\ \emph {et~al.}(2012)\citenamefont
  {Bardarson}, \citenamefont {Pollmann},\ and\ \citenamefont
  {Moore}}]{bardarson_unbounded_2012}%
  \BibitemOpen
  \bibfield  {author} {\bibinfo {author} {\bibfnamefont {J.~H.}\ \bibnamefont
  {Bardarson}}, \bibinfo {author} {\bibfnamefont {F.}~\bibnamefont {Pollmann}},
  \ and\ \bibinfo {author} {\bibfnamefont {J.~E.}\ \bibnamefont {Moore}},\
  }\href {\doibase 10.1103/PhysRevLett.109.017202} {\bibfield  {journal}
  {\bibinfo  {journal} {Phys. Rev. Lett.}\ }\textbf {\bibinfo {volume} {109}},\
  \bibinfo {pages} {017202} (\bibinfo {year} {2012})}\BibitemShut {NoStop}%
\bibitem [{\citenamefont {Luitz}\ \emph {et~al.}(2015)\citenamefont {Luitz},
  \citenamefont {Laflorencie},\ and\ \citenamefont
  {Alet}}]{luitz_many-body_2015}%
  \BibitemOpen
  \bibfield  {author} {\bibinfo {author} {\bibfnamefont {D.~J.}\ \bibnamefont
  {Luitz}}, \bibinfo {author} {\bibfnamefont {N.}~\bibnamefont {Laflorencie}},
  \ and\ \bibinfo {author} {\bibfnamefont {F.}~\bibnamefont {Alet}},\ }\href
  {\doibase 10.1103/PhysRevB.91.081103} {\bibfield  {journal} {\bibinfo
  {journal} {Phys. Rev. B}\ }\textbf {\bibinfo {volume} {91}},\ \bibinfo
  {pages} {081103} (\bibinfo {year} {2015})}\BibitemShut {NoStop}%
\bibitem [{\citenamefont {Alet}\ and\ \citenamefont
  {Laflorencie}(2018)}]{alet_many-body_2018}%
  \BibitemOpen
  \bibfield  {author} {\bibinfo {author} {\bibfnamefont {F.}~\bibnamefont
  {Alet}}\ and\ \bibinfo {author} {\bibfnamefont {N.}~\bibnamefont
  {Laflorencie}},\ }\href {\doibase 10.1016/j.crhy.2018.03.003} {\bibfield
  {journal} {\bibinfo  {journal} {Comptes Rendus Physique}\ }\textbf {\bibinfo
  {volume} {19}},\ \bibinfo {pages} {498} (\bibinfo {year} {2018})}\BibitemShut
  {NoStop}%
\bibitem [{\citenamefont {Abanin}\ \emph {et~al.}(2019)\citenamefont {Abanin},
  \citenamefont {Altman}, \citenamefont {Bloch},\ and\ \citenamefont
  {Serbyn}}]{abanin_many-body_2019}%
  \BibitemOpen
  \bibfield  {author} {\bibinfo {author} {\bibfnamefont {D.~A.}\ \bibnamefont
  {Abanin}}, \bibinfo {author} {\bibfnamefont {E.}~\bibnamefont {Altman}},
  \bibinfo {author} {\bibfnamefont {I.}~\bibnamefont {Bloch}}, \ and\ \bibinfo
  {author} {\bibfnamefont {M.}~\bibnamefont {Serbyn}},\ }\href {\doibase
  10.1103/RevModPhys.91.021001} {\bibfield  {journal} {\bibinfo  {journal}
  {Rev. Mod. Phys.}\ }\textbf {\bibinfo {volume} {91}},\ \bibinfo {pages}
  {021001} (\bibinfo {year} {2019})}\BibitemShut {NoStop}%
\bibitem [{\citenamefont {Doggen}\ \emph {et~al.}(2018)\citenamefont {Doggen},
  \citenamefont {Schindler}, \citenamefont {Tikhonov}, \citenamefont {Mirlin},
  \citenamefont {Neupert}, \citenamefont {Polyakov},\ and\ \citenamefont
  {Gornyi}}]{doggen_many-body_2018}%
  \BibitemOpen
  \bibfield  {author} {\bibinfo {author} {\bibfnamefont {E.~V.~H.}\
  \bibnamefont {Doggen}}, \bibinfo {author} {\bibfnamefont {F.}~\bibnamefont
  {Schindler}}, \bibinfo {author} {\bibfnamefont {K.~S.}\ \bibnamefont
  {Tikhonov}}, \bibinfo {author} {\bibfnamefont {A.~D.}\ \bibnamefont
  {Mirlin}}, \bibinfo {author} {\bibfnamefont {T.}~\bibnamefont {Neupert}},
  \bibinfo {author} {\bibfnamefont {D.~G.}\ \bibnamefont {Polyakov}}, \ and\
  \bibinfo {author} {\bibfnamefont {I.~V.}\ \bibnamefont {Gornyi}},\ }\href
  {\doibase 10.1103/PhysRevB.98.174202} {\bibfield  {journal} {\bibinfo
  {journal} {Phys. Rev. B}\ }\textbf {\bibinfo {volume} {98}},\ \bibinfo
  {pages} {174202} (\bibinfo {year} {2018})}\BibitemShut {NoStop}%
\bibitem [{\citenamefont {Chanda}\ \emph {et~al.}(2020)\citenamefont {Chanda},
  \citenamefont {Sierant},\ and\ \citenamefont
  {Zakrzewski}}]{chanda_time_2020}%
  \BibitemOpen
  \bibfield  {author} {\bibinfo {author} {\bibfnamefont {T.}~\bibnamefont
  {Chanda}}, \bibinfo {author} {\bibfnamefont {P.}~\bibnamefont {Sierant}}, \
  and\ \bibinfo {author} {\bibfnamefont {J.}~\bibnamefont {Zakrzewski}},\
  }\href {\doibase 10.1103/PhysRevB.101.035148} {\bibfield  {journal} {\bibinfo
   {journal} {Phys. Rev. B}\ }\textbf {\bibinfo {volume} {101}},\ \bibinfo
  {pages} {035148} (\bibinfo {year} {2020})}\BibitemShut {NoStop}%
\bibitem [{\citenamefont {Sierant}\ \emph {et~al.}(2020)\citenamefont
  {Sierant}, \citenamefont {Delande},\ and\ \citenamefont
  {Zakrzewski}}]{sierant_thouless_2020}%
  \BibitemOpen
  \bibfield  {author} {\bibinfo {author} {\bibfnamefont {P.}~\bibnamefont
  {Sierant}}, \bibinfo {author} {\bibfnamefont {D.}~\bibnamefont {Delande}}, \
  and\ \bibinfo {author} {\bibfnamefont {J.}~\bibnamefont {Zakrzewski}},\
  }\href {\doibase 10.1103/PhysRevLett.124.186601} {\bibfield  {journal}
  {\bibinfo  {journal} {Phys. Rev. Lett.}\ }\textbf {\bibinfo {volume} {124}},\
  \bibinfo {pages} {186601} (\bibinfo {year} {2020})}\BibitemShut {NoStop}%
\bibitem [{\citenamefont {Abanin}\ \emph {et~al.}(2021)\citenamefont {Abanin},
  \citenamefont {Bardarson}, \citenamefont {De~Tomasi}, \citenamefont
  {Gopalakrishnan}, \citenamefont {Khemani}, \citenamefont {Parameswaran},
  \citenamefont {Pollmann}, \citenamefont {Potter}, \citenamefont {Serbyn},\
  and\ \citenamefont {Vasseur}}]{abanin_distinguishing_2021}%
  \BibitemOpen
  \bibfield  {author} {\bibinfo {author} {\bibfnamefont {D.~A.}\ \bibnamefont
  {Abanin}}, \bibinfo {author} {\bibfnamefont {J.~H.}\ \bibnamefont
  {Bardarson}}, \bibinfo {author} {\bibfnamefont {G.}~\bibnamefont
  {De~Tomasi}}, \bibinfo {author} {\bibfnamefont {S.}~\bibnamefont
  {Gopalakrishnan}}, \bibinfo {author} {\bibfnamefont {V.}~\bibnamefont
  {Khemani}}, \bibinfo {author} {\bibfnamefont {S.~A.}\ \bibnamefont
  {Parameswaran}}, \bibinfo {author} {\bibfnamefont {F.}~\bibnamefont
  {Pollmann}}, \bibinfo {author} {\bibfnamefont {A.~C.}\ \bibnamefont
  {Potter}}, \bibinfo {author} {\bibfnamefont {M.}~\bibnamefont {Serbyn}}, \
  and\ \bibinfo {author} {\bibfnamefont {R.}~\bibnamefont {Vasseur}},\ }\href
  {\doibase 10.1016/j.aop.2021.168415} {\bibfield  {journal} {\bibinfo
  {journal} {Annals of Physics}\ }\textbf {\bibinfo {volume} {427}},\ \bibinfo
  {pages} {168415} (\bibinfo {year} {2021})}\BibitemShut {NoStop}%
\bibitem [{\citenamefont {Schreiber}\ \emph {et~al.}(2015)\citenamefont
  {Schreiber}, \citenamefont {Hodgman}, \citenamefont {Bordia}, \citenamefont
  {L{\"u}schen}, \citenamefont {Fischer}, \citenamefont {Vosk}, \citenamefont
  {Altman}, \citenamefont {Schneider},\ and\ \citenamefont
  {Bloch}}]{schreiber_observation_2015}%
  \BibitemOpen
  \bibfield  {author} {\bibinfo {author} {\bibfnamefont {M.}~\bibnamefont
  {Schreiber}}, \bibinfo {author} {\bibfnamefont {S.~S.}\ \bibnamefont
  {Hodgman}}, \bibinfo {author} {\bibfnamefont {P.}~\bibnamefont {Bordia}},
  \bibinfo {author} {\bibfnamefont {H.~P.}\ \bibnamefont {L{\"u}schen}},
  \bibinfo {author} {\bibfnamefont {M.~H.}\ \bibnamefont {Fischer}}, \bibinfo
  {author} {\bibfnamefont {R.}~\bibnamefont {Vosk}}, \bibinfo {author}
  {\bibfnamefont {E.}~\bibnamefont {Altman}}, \bibinfo {author} {\bibfnamefont
  {U.}~\bibnamefont {Schneider}}, \ and\ \bibinfo {author} {\bibfnamefont
  {I.}~\bibnamefont {Bloch}},\ }\href {\doibase 10.1126/science.aaa7432}
  {\bibfield  {journal} {\bibinfo  {journal} {Science}\ }\textbf {\bibinfo
  {volume} {349}},\ \bibinfo {pages} {842} (\bibinfo {year}
  {2015})}\BibitemShut {NoStop}%
\bibitem [{\citenamefont {Smith}\ \emph {et~al.}(2016)\citenamefont {Smith},
  \citenamefont {Lee}, \citenamefont {Richerme}, \citenamefont {Neyenhuis},
  \citenamefont {Hess}, \citenamefont {Hauke}, \citenamefont {Heyl},
  \citenamefont {Huse},\ and\ \citenamefont {Monroe}}]{smith_many-body_2016}%
  \BibitemOpen
  \bibfield  {author} {\bibinfo {author} {\bibfnamefont {J.}~\bibnamefont
  {Smith}}, \bibinfo {author} {\bibfnamefont {A.}~\bibnamefont {Lee}}, \bibinfo
  {author} {\bibfnamefont {P.}~\bibnamefont {Richerme}}, \bibinfo {author}
  {\bibfnamefont {B.}~\bibnamefont {Neyenhuis}}, \bibinfo {author}
  {\bibfnamefont {P.~W.}\ \bibnamefont {Hess}}, \bibinfo {author}
  {\bibfnamefont {P.}~\bibnamefont {Hauke}}, \bibinfo {author} {\bibfnamefont
  {M.}~\bibnamefont {Heyl}}, \bibinfo {author} {\bibfnamefont {D.~A.}\
  \bibnamefont {Huse}}, \ and\ \bibinfo {author} {\bibfnamefont
  {C.}~\bibnamefont {Monroe}},\ }\href {\doibase 10.1038/nphys3783} {\bibfield
  {journal} {\bibinfo  {journal} {Nature Physics}\ }\textbf {\bibinfo {volume}
  {12}},\ \bibinfo {pages} {907} (\bibinfo {year} {2016})}\BibitemShut
  {NoStop}%
\bibitem [{\citenamefont {Choi}\ \emph {et~al.}(2016)\citenamefont {Choi},
  \citenamefont {Hild}, \citenamefont {Zeiher}, \citenamefont {Schau{\ss}},
  \citenamefont {Rubio-Abadal}, \citenamefont {Yefsah}, \citenamefont
  {Khemani}, \citenamefont {Huse}, \citenamefont {Bloch},\ and\ \citenamefont
  {Gross}}]{choi_exploring_2016}%
  \BibitemOpen
  \bibfield  {author} {\bibinfo {author} {\bibfnamefont {J.-y.}\ \bibnamefont
  {Choi}}, \bibinfo {author} {\bibfnamefont {S.}~\bibnamefont {Hild}}, \bibinfo
  {author} {\bibfnamefont {J.}~\bibnamefont {Zeiher}}, \bibinfo {author}
  {\bibfnamefont {P.}~\bibnamefont {Schau{\ss}}}, \bibinfo {author}
  {\bibfnamefont {A.}~\bibnamefont {Rubio-Abadal}}, \bibinfo {author}
  {\bibfnamefont {T.}~\bibnamefont {Yefsah}}, \bibinfo {author} {\bibfnamefont
  {V.}~\bibnamefont {Khemani}}, \bibinfo {author} {\bibfnamefont {D.~A.}\
  \bibnamefont {Huse}}, \bibinfo {author} {\bibfnamefont {I.}~\bibnamefont
  {Bloch}}, \ and\ \bibinfo {author} {\bibfnamefont {C.}~\bibnamefont
  {Gross}},\ }\href {\doibase 10.1126/science.aaf8834} {\bibfield  {journal}
  {\bibinfo  {journal} {Science}\ }\textbf {\bibinfo {volume} {352}},\ \bibinfo
  {pages} {1547} (\bibinfo {year} {2016})}\BibitemShut {NoStop}%
\bibitem [{\citenamefont {Roushan~{\it{et
  al.}}}(2017)}]{roushan_spectroscopic_2017}%
  \BibitemOpen
  \bibfield  {author} {\bibinfo {author} {\bibfnamefont {P.}~\bibnamefont
  {Roushan~{\it{et al.}}}},\ }\href {\doibase 10.1126/science.aao1401}
  {\bibfield  {journal} {\bibinfo  {journal} {Science}\ }\textbf {\bibinfo
  {volume} {358}},\ \bibinfo {pages} {1175} (\bibinfo {year}
  {2017})}\BibitemShut {NoStop}%
\bibitem [{\citenamefont {Imbrie}(2016)}]{imbrie_many-body_2016}%
  \BibitemOpen
  \bibfield  {author} {\bibinfo {author} {\bibfnamefont {J.~Z.}\ \bibnamefont
  {Imbrie}},\ }\href {\doibase 10.1007/s10955-016-1508-x} {\bibfield  {journal}
  {\bibinfo  {journal} {J Stat Phys}\ }\textbf {\bibinfo {volume} {163}},\
  \bibinfo {pages} {998} (\bibinfo {year} {2016})}\BibitemShut {NoStop}%
\bibitem [{\citenamefont {Pekker}\ \emph {et~al.}(2014)\citenamefont {Pekker},
  \citenamefont {Refael}, \citenamefont {Altman}, \citenamefont {Demler},\ and\
  \citenamefont {Oganesyan}}]{pekker_hilbert-glass_2014}%
  \BibitemOpen
  \bibfield  {author} {\bibinfo {author} {\bibfnamefont {D.}~\bibnamefont
  {Pekker}}, \bibinfo {author} {\bibfnamefont {G.}~\bibnamefont {Refael}},
  \bibinfo {author} {\bibfnamefont {E.}~\bibnamefont {Altman}}, \bibinfo
  {author} {\bibfnamefont {E.}~\bibnamefont {Demler}}, \ and\ \bibinfo {author}
  {\bibfnamefont {V.}~\bibnamefont {Oganesyan}},\ }\href {\doibase
  10.1103/PhysRevX.4.011052} {\bibfield  {journal} {\bibinfo  {journal} {Phys.
  Rev. X}\ }\textbf {\bibinfo {volume} {4}},\ \bibinfo {pages} {011052}
  (\bibinfo {year} {2014})}\BibitemShut {NoStop}%
\bibitem [{\citenamefont {Kj{\"a}ll}\ \emph {et~al.}(2014)\citenamefont
  {Kj{\"a}ll}, \citenamefont {Bardarson},\ and\ \citenamefont
  {Pollmann}}]{kjall_many-body_2014}%
  \BibitemOpen
  \bibfield  {author} {\bibinfo {author} {\bibfnamefont {J.~A.}\ \bibnamefont
  {Kj{\"a}ll}}, \bibinfo {author} {\bibfnamefont {J.~H.}\ \bibnamefont
  {Bardarson}}, \ and\ \bibinfo {author} {\bibfnamefont {F.}~\bibnamefont
  {Pollmann}},\ }\href {\doibase 10.1103/PhysRevLett.113.107204} {\bibfield
  {journal} {\bibinfo  {journal} {Phys. Rev. Lett.}\ }\textbf {\bibinfo
  {volume} {113}},\ \bibinfo {pages} {107204} (\bibinfo {year}
  {2014})}\BibitemShut {NoStop}%
\bibitem [{\citenamefont {Sahay}\ \emph {et~al.}(2021)\citenamefont {Sahay},
  \citenamefont {Machado}, \citenamefont {Ye}, \citenamefont {Laumann},\ and\
  \citenamefont {Yao}}]{sahay_emergent_2021}%
  \BibitemOpen
  \bibfield  {author} {\bibinfo {author} {\bibfnamefont {R.}~\bibnamefont
  {Sahay}}, \bibinfo {author} {\bibfnamefont {F.}~\bibnamefont {Machado}},
  \bibinfo {author} {\bibfnamefont {B.}~\bibnamefont {Ye}}, \bibinfo {author}
  {\bibfnamefont {C.~R.}\ \bibnamefont {Laumann}}, \ and\ \bibinfo {author}
  {\bibfnamefont {N.~Y.}\ \bibnamefont {Yao}},\ }\href {\doibase
  10.1103/PhysRevLett.126.100604} {\bibfield  {journal} {\bibinfo  {journal}
  {Phys. Rev. Lett.}\ }\textbf {\bibinfo {volume} {126}},\ \bibinfo {pages}
  {100604} (\bibinfo {year} {2021})}\BibitemShut {NoStop}%
\bibitem [{\citenamefont {Moudgalya}\ \emph {et~al.}(2020)\citenamefont
  {Moudgalya}, \citenamefont {Huse},\ and\ \citenamefont
  {Khemani}}]{moudgalya_perturbative_2020}%
  \BibitemOpen
  \bibfield  {author} {\bibinfo {author} {\bibfnamefont {S.}~\bibnamefont
  {Moudgalya}}, \bibinfo {author} {\bibfnamefont {D.~A.}\ \bibnamefont {Huse}},
  \ and\ \bibinfo {author} {\bibfnamefont {V.}~\bibnamefont {Khemani}},\ }\href
  {http://arxiv.org/abs/2008.09113} {\bibfield  {journal} {\bibinfo  {journal}
  {arXiv:2008.09113}\ } (\bibinfo {year} {2020})}\BibitemShut {NoStop}%
\bibitem [{\citenamefont {Wahl}\ \emph {et~al.}(2021)\citenamefont {Wahl},
  \citenamefont {Venn},\ and\ \citenamefont {B{\'e}ri}}]{wahl2021local}%
  \BibitemOpen
  \bibfield  {author} {\bibinfo {author} {\bibfnamefont {T.~B.}\ \bibnamefont
  {Wahl}}, \bibinfo {author} {\bibfnamefont {F.}~\bibnamefont {Venn}}, \ and\
  \bibinfo {author} {\bibfnamefont {B.}~\bibnamefont {B{\'e}ri}},\ }\href
  {https://arxiv.org/abs/2111.11543} {} (\bibinfo {year} {2021}),\ \Eprint
  {http://arxiv.org/abs/2111.11543} {arXiv:2111.11543} \BibitemShut {NoStop}%
\bibitem [{\citenamefont {Huse}\ \emph {et~al.}(2013)\citenamefont {Huse},
  \citenamefont {Nandkishore}, \citenamefont {Oganesyan}, \citenamefont {Pal},\
  and\ \citenamefont {Sondhi}}]{huse_localization-protected_2013}%
  \BibitemOpen
  \bibfield  {author} {\bibinfo {author} {\bibfnamefont {D.~A.}\ \bibnamefont
  {Huse}}, \bibinfo {author} {\bibfnamefont {R.}~\bibnamefont {Nandkishore}},
  \bibinfo {author} {\bibfnamefont {V.}~\bibnamefont {Oganesyan}}, \bibinfo
  {author} {\bibfnamefont {A.}~\bibnamefont {Pal}}, \ and\ \bibinfo {author}
  {\bibfnamefont {S.~L.}\ \bibnamefont {Sondhi}},\ }\href {\doibase
  10.1103/PhysRevB.88.014206} {\bibfield  {journal} {\bibinfo  {journal} {Phys.
  Rev. B}\ }\textbf {\bibinfo {volume} {88}},\ \bibinfo {pages} {014206}
  (\bibinfo {year} {2013})}\BibitemShut {NoStop}%
\bibitem [{\citenamefont {{Fendley}}(2012)}]{fendley_parafermionic_2012}%
  \BibitemOpen
  \bibfield  {author} {\bibinfo {author} {\bibfnamefont {P.}~\bibnamefont
  {{Fendley}}},\ }\href {\doibase 10.1088/1742-5468/2012/11/P11020} {\bibfield
  {journal} {\bibinfo  {journal} {Journal of Statistical Mechanics: Theory and
  Experiment}\ }\textbf {\bibinfo {volume} {2012}},\ \bibinfo {pages} {11020}
  (\bibinfo {year} {2012})}\BibitemShut {NoStop}%
\bibitem [{\citenamefont {Fisher}(1995)}]{fisher_critical_1995}%
  \BibitemOpen
  \bibfield  {author} {\bibinfo {author} {\bibfnamefont {D.~S.}\ \bibnamefont
  {Fisher}},\ }\href {\doibase 10.1103/PhysRevB.51.6411} {\bibfield  {journal}
  {\bibinfo  {journal} {Phys. Rev. B}\ }\textbf {\bibinfo {volume} {51}},\
  \bibinfo {pages} {6411} (\bibinfo {year} {1995})}\BibitemShut {NoStop}%
\bibitem [{\citenamefont {De~Roeck}\ and\ \citenamefont
  {Huveneers}(2017)}]{de_roeck_stability_2017}%
  \BibitemOpen
  \bibfield  {author} {\bibinfo {author} {\bibfnamefont {W.}~\bibnamefont
  {De~Roeck}}\ and\ \bibinfo {author} {\bibfnamefont {F.}~\bibnamefont
  {Huveneers}},\ }\href {\doibase 10.1103/PhysRevB.95.155129} {\bibfield
  {journal} {\bibinfo  {journal} {Phys. Rev. B}\ }\textbf {\bibinfo {volume}
  {95}},\ \bibinfo {pages} {155129} (\bibinfo {year} {2017})}\BibitemShut
  {NoStop}%
\bibitem [{\citenamefont {Bahri}\ \emph {et~al.}(2015)\citenamefont {Bahri},
  \citenamefont {Vosk}, \citenamefont {Altman},\ and\ \citenamefont
  {Vishwanath}}]{bahri_localization_2015}%
  \BibitemOpen
  \bibfield  {author} {\bibinfo {author} {\bibfnamefont {Y.}~\bibnamefont
  {Bahri}}, \bibinfo {author} {\bibfnamefont {R.}~\bibnamefont {Vosk}},
  \bibinfo {author} {\bibfnamefont {E.}~\bibnamefont {Altman}}, \ and\ \bibinfo
  {author} {\bibfnamefont {A.}~\bibnamefont {Vishwanath}},\ }\href {\doibase
  10.1038/ncomms8341} {\bibfield  {journal} {\bibinfo  {journal} {Nature
  Communications}\ }\textbf {\bibinfo {volume} {6}},\ \bibinfo {pages} {7341}
  (\bibinfo {year} {2015})}\BibitemShut {NoStop}%
\bibitem [{\citenamefont {Else}\ \emph {et~al.}(2017)\citenamefont {Else},
  \citenamefont {Fendley}, \citenamefont {Kemp},\ and\ \citenamefont
  {Nayak}}]{PhysRevX.7.041062}%
  \BibitemOpen
  \bibfield  {author} {\bibinfo {author} {\bibfnamefont {D.~V.}\ \bibnamefont
  {Else}}, \bibinfo {author} {\bibfnamefont {P.}~\bibnamefont {Fendley}},
  \bibinfo {author} {\bibfnamefont {J.}~\bibnamefont {Kemp}}, \ and\ \bibinfo
  {author} {\bibfnamefont {C.}~\bibnamefont {Nayak}},\ }\href {\doibase
  10.1103/PhysRevX.7.041062} {\bibfield  {journal} {\bibinfo  {journal} {Phys.
  Rev. X}\ }\textbf {\bibinfo {volume} {7}},\ \bibinfo {pages} {041062}
  (\bibinfo {year} {2017})}\BibitemShut {NoStop}%
\bibitem [{\citenamefont {Kramers}\ and\ \citenamefont
  {Wannier}(1941)}]{kramers_statistics_1941}%
  \BibitemOpen
  \bibfield  {author} {\bibinfo {author} {\bibfnamefont {H.~A.}\ \bibnamefont
  {Kramers}}\ and\ \bibinfo {author} {\bibfnamefont {G.~H.}\ \bibnamefont
  {Wannier}},\ }\href {\doibase 10.1103/PhysRev.60.252} {\bibfield  {journal}
  {\bibinfo  {journal} {Phys. Rev.}\ }\textbf {\bibinfo {volume} {60}},\
  \bibinfo {pages} {252} (\bibinfo {year} {1941})}\BibitemShut {NoStop}%
\bibitem [{noteSZM()}]{noteSZM}%
  \BibitemOpen
  \bibinfo {note} {The SZM operator pairs states in different symmetry (here
  the parity) sectors, leading to identical spectra up to exponentially small
  finite-size corrections, see also \cite{Fendley_2016}.}\BibitemShut {Stop}%
\bibitem [{\citenamefont {Rahmani}\ \emph {et~al.}(2015)\citenamefont
  {Rahmani}, \citenamefont {Zhu}, \citenamefont {Franz},\ and\ \citenamefont
  {Affleck}}]{rahmani_phase_2015}%
  \BibitemOpen
  \bibfield  {author} {\bibinfo {author} {\bibfnamefont {A.}~\bibnamefont
  {Rahmani}}, \bibinfo {author} {\bibfnamefont {X.}~\bibnamefont {Zhu}},
  \bibinfo {author} {\bibfnamefont {M.}~\bibnamefont {Franz}}, \ and\ \bibinfo
  {author} {\bibfnamefont {I.}~\bibnamefont {Affleck}},\ }\href {\doibase
  10.1103/PhysRevB.92.235123} {\bibfield  {journal} {\bibinfo  {journal} {Phys.
  Rev. B}\ }\textbf {\bibinfo {volume} {92}},\ \bibinfo {pages} {235123}
  (\bibinfo {year} {2015})}\BibitemShut {NoStop}%
\bibitem [{\citenamefont {Lobos}\ \emph {et~al.}(2012)\citenamefont {Lobos},
  \citenamefont {Lutchyn},\ and\ \citenamefont
  {Das~Sarma}}]{lobos_interplay_2012}%
  \BibitemOpen
  \bibfield  {author} {\bibinfo {author} {\bibfnamefont {A.~M.}\ \bibnamefont
  {Lobos}}, \bibinfo {author} {\bibfnamefont {R.~M.}\ \bibnamefont {Lutchyn}},
  \ and\ \bibinfo {author} {\bibfnamefont {S.}~\bibnamefont {Das~Sarma}},\
  }\href {\doibase 10.1103/PhysRevLett.109.146403} {\bibfield  {journal}
  {\bibinfo  {journal} {Phys. Rev. Lett.}\ }\textbf {\bibinfo {volume} {109}},\
  \bibinfo {pages} {146403} (\bibinfo {year} {2012})}\BibitemShut {NoStop}%
\bibitem [{\citenamefont {Cr{\'e}pin}\ \emph {et~al.}(2014)\citenamefont
  {Cr{\'e}pin}, \citenamefont {Zar{\'a}nd},\ and\ \citenamefont
  {Simon}}]{crepin_nonperturbative_2014}%
  \BibitemOpen
  \bibfield  {author} {\bibinfo {author} {\bibfnamefont {F.}~\bibnamefont
  {Cr{\'e}pin}}, \bibinfo {author} {\bibfnamefont {G.}~\bibnamefont
  {Zar{\'a}nd}}, \ and\ \bibinfo {author} {\bibfnamefont {P.}~\bibnamefont
  {Simon}},\ }\href {\doibase 10.1103/PhysRevB.90.121407} {\bibfield  {journal}
  {\bibinfo  {journal} {Phys. Rev. B}\ }\textbf {\bibinfo {volume} {90}},\
  \bibinfo {pages} {121407} (\bibinfo {year} {2014})}\BibitemShut {NoStop}%
\bibitem [{\citenamefont {Gergs}\ \emph {et~al.}(2016)\citenamefont {Gergs},
  \citenamefont {Fritz},\ and\ \citenamefont
  {Schuricht}}]{gergs_topological_2016}%
  \BibitemOpen
  \bibfield  {author} {\bibinfo {author} {\bibfnamefont {N.~M.}\ \bibnamefont
  {Gergs}}, \bibinfo {author} {\bibfnamefont {L.}~\bibnamefont {Fritz}}, \ and\
  \bibinfo {author} {\bibfnamefont {D.}~\bibnamefont {Schuricht}},\ }\href
  {\doibase 10.1103/PhysRevB.93.075129} {\bibfield  {journal} {\bibinfo
  {journal} {Phys. Rev. B}\ }\textbf {\bibinfo {volume} {93}},\ \bibinfo
  {pages} {075129} (\bibinfo {year} {2016})}\BibitemShut {NoStop}%
\bibitem [{\citenamefont {Karcher}\ \emph {et~al.}(2019)\citenamefont
  {Karcher}, \citenamefont {Sonner},\ and\ \citenamefont
  {Mirlin}}]{karcher_disorder_2019}%
  \BibitemOpen
  \bibfield  {author} {\bibinfo {author} {\bibfnamefont {J.~F.}\ \bibnamefont
  {Karcher}}, \bibinfo {author} {\bibfnamefont {M.}~\bibnamefont {Sonner}}, \
  and\ \bibinfo {author} {\bibfnamefont {A.~D.}\ \bibnamefont {Mirlin}},\
  }\href {\doibase 10.1103/PhysRevB.100.134207} {\bibfield  {journal} {\bibinfo
   {journal} {Phys. Rev. B}\ }\textbf {\bibinfo {volume} {100}},\ \bibinfo
  {pages} {134207} (\bibinfo {year} {2019})}\BibitemShut {NoStop}%
\bibitem [{\citenamefont {Kitaev}(2001)}]{kitaev_unpaired_2001}%
  \BibitemOpen
  \bibfield  {author} {\bibinfo {author} {\bibfnamefont {A.~Y.}\ \bibnamefont
  {Kitaev}},\ }\href {\doibase 10.1070/1063-7869/44/10S/S29} {\bibfield
  {journal} {\bibinfo  {journal} {Phys.-Usp.}\ }\textbf {\bibinfo {volume}
  {44}},\ \bibinfo {pages} {131} (\bibinfo {year} {2001})}\BibitemShut
  {NoStop}%
\bibitem [{\citenamefont {Lin}\ \emph {et~al.}(2018)\citenamefont {Lin},
  \citenamefont {Sbierski}, \citenamefont {Dorfner}, \citenamefont {Karrasch},\
  and\ \citenamefont {Heidrich-Meisner}}]{lin_many-body_2018}%
  \BibitemOpen
  \bibfield  {author} {\bibinfo {author} {\bibfnamefont {S.-H.}\ \bibnamefont
  {Lin}}, \bibinfo {author} {\bibfnamefont {B.}~\bibnamefont {Sbierski}},
  \bibinfo {author} {\bibfnamefont {F.}~\bibnamefont {Dorfner}}, \bibinfo
  {author} {\bibfnamefont {C.}~\bibnamefont {Karrasch}}, \ and\ \bibinfo
  {author} {\bibfnamefont {F.}~\bibnamefont {Heidrich-Meisner}},\ }\href
  {\doibase 10.21468/SciPostPhys.4.1.002} {\bibfield  {journal} {\bibinfo
  {journal} {SciPost Physics}\ }\textbf {\bibinfo {volume} {4}},\ \bibinfo
  {pages} {002} (\bibinfo {year} {2018})}\BibitemShut {NoStop}%
\bibitem [{\citenamefont {Pietracaprina}\ \emph {et~al.}(2018)\citenamefont
  {Pietracaprina}, \citenamefont {Mac{\'e}}, \citenamefont {Luitz},\ and\
  \citenamefont {Alet}}]{pietracaprina_shift-invert_2018}%
  \BibitemOpen
  \bibfield  {author} {\bibinfo {author} {\bibfnamefont {F.}~\bibnamefont
  {Pietracaprina}}, \bibinfo {author} {\bibfnamefont {N.}~\bibnamefont
  {Mac{\'e}}}, \bibinfo {author} {\bibfnamefont {D.~J.}\ \bibnamefont {Luitz}},
  \ and\ \bibinfo {author} {\bibfnamefont {F.}~\bibnamefont {Alet}},\ }\href
  {\doibase 10.21468/SciPostPhys.5.5.045} {\bibfield  {journal} {\bibinfo
  {journal} {SciPost Physics}\ }\textbf {\bibinfo {volume} {5}},\ \bibinfo
  {pages} {045} (\bibinfo {year} {2018})}\BibitemShut {NoStop}%
\bibitem [{Note1()}]{Note1}%
  \BibitemOpen
  \bibinfo {note} {This corresponds to 32 Majorana fermions, and Hilbert spaces
  of maximum size 65\protect \tmspace +\thinmuskip {.1667em}536.}\BibitemShut
  {Stop}%
\bibitem [{Note2()}]{Note2}%
  \BibitemOpen
  \bibinfo {note} {For odd system sizes $L=2p+1$ we cut at the bond
  $x=p$.}\BibitemShut {Stop}%
\bibitem [{\citenamefont {Oganesyan}\ and\ \citenamefont
  {Huse}(2007)}]{oganesyan_localization_2007}%
  \BibitemOpen
  \bibfield  {author} {\bibinfo {author} {\bibfnamefont {V.}~\bibnamefont
  {Oganesyan}}\ and\ \bibinfo {author} {\bibfnamefont {D.~A.}\ \bibnamefont
  {Huse}},\ }\href {\doibase 10.1103/PhysRevB.75.155111} {\bibfield  {journal}
  {\bibinfo  {journal} {Phys. Rev. B}\ }\textbf {\bibinfo {volume} {75}},\
  \bibinfo {pages} {155111} (\bibinfo {year} {2007})}\BibitemShut {NoStop}%
\bibitem [{\citenamefont {Atas}\ \emph {et~al.}(2013)\citenamefont {Atas},
  \citenamefont {Bogomolny}, \citenamefont {Giraud},\ and\ \citenamefont
  {Roux}}]{atas_distribution_2013}%
  \BibitemOpen
  \bibfield  {author} {\bibinfo {author} {\bibfnamefont {Y.~Y.}\ \bibnamefont
  {Atas}}, \bibinfo {author} {\bibfnamefont {E.}~\bibnamefont {Bogomolny}},
  \bibinfo {author} {\bibfnamefont {O.}~\bibnamefont {Giraud}}, \ and\ \bibinfo
  {author} {\bibfnamefont {G.}~\bibnamefont {Roux}},\ }\href {\doibase
  10.1103/PhysRevLett.110.084101} {\bibfield  {journal} {\bibinfo  {journal}
  {Phys. Rev. Lett.}\ }\textbf {\bibinfo {volume} {110}},\ \bibinfo {pages}
  {084101} (\bibinfo {year} {2013})}\BibitemShut {NoStop}%
\bibitem [{sm()}]{sm}%
  \BibitemOpen
  \bibinfo {note} {See supplemental Material.}\BibitemShut {Stop}%
\bibitem [{\citenamefont {Laflorencie}(2022)}]{laflo_2022}%
  \BibitemOpen
  \bibfield  {author} {\bibinfo {author} {\bibfnamefont {N.}~\bibnamefont
  {Laflorencie}},\ }\href@noop {} {} (\bibinfo {year}
  {unpublished~2022})\BibitemShut {NoStop}%
\bibitem [{\citenamefont {Fendley}(2016)}]{Fendley_2016}%
  \BibitemOpen
  \bibfield  {author} {\bibinfo {author} {\bibfnamefont {P.}~\bibnamefont
  {Fendley}},\ }\href {\doibase 10.1088/1751-8113/49/30/30lt01} {\bibfield
  {journal} {\bibinfo  {journal} {Journal of Physics A: Mathematical and
  Theoretical}\ }\textbf {\bibinfo {volume} {49}},\ \bibinfo {pages} {30LT01}
  (\bibinfo {year} {2016})}\BibitemShut {NoStop}%
\bibitem [{\citenamefont {Kemp}\ \emph {et~al.}(2017)\citenamefont {Kemp},
  \citenamefont {Yao}, \citenamefont {Laumann},\ and\ \citenamefont
  {Fendley}}]{Kemp_2017}%
  \BibitemOpen
  \bibfield  {author} {\bibinfo {author} {\bibfnamefont {J.}~\bibnamefont
  {Kemp}}, \bibinfo {author} {\bibfnamefont {N.~Y.}\ \bibnamefont {Yao}},
  \bibinfo {author} {\bibfnamefont {C.~R.}\ \bibnamefont {Laumann}}, \ and\
  \bibinfo {author} {\bibfnamefont {P.}~\bibnamefont {Fendley}},\ }\href
  {\doibase 10.1088/1742-5468/aa73f0} {\bibfield  {journal} {\bibinfo
  {journal} {Journal of Statistical Mechanics: Theory and Experiment}\ }\textbf
  {\bibinfo {volume} {2017}},\ \bibinfo {pages} {063105} (\bibinfo {year}
  {2017})}\BibitemShut {NoStop}%
\bibitem [{\citenamefont {Vasseur}\ \emph {et~al.}(2016)\citenamefont
  {Vasseur}, \citenamefont {Friedman}, \citenamefont {Parameswaran},\ and\
  \citenamefont {Potter}}]{vasseur_particle-hole_2016}%
  \BibitemOpen
  \bibfield  {author} {\bibinfo {author} {\bibfnamefont {R.}~\bibnamefont
  {Vasseur}}, \bibinfo {author} {\bibfnamefont {A.~J.}\ \bibnamefont
  {Friedman}}, \bibinfo {author} {\bibfnamefont {S.~A.}\ \bibnamefont
  {Parameswaran}}, \ and\ \bibinfo {author} {\bibfnamefont {A.~C.}\
  \bibnamefont {Potter}},\ }\href {\doibase 10.1103/PhysRevB.93.134207}
  {\bibfield  {journal} {\bibinfo  {journal} {Phys. Rev. B}\ }\textbf {\bibinfo
  {volume} {93}},\ \bibinfo {pages} {134207} (\bibinfo {year}
  {2016})}\BibitemShut {NoStop}%
\bibitem [{\citenamefont {Giraud}\ \emph {et~al.}(2020)\citenamefont {Giraud},
  \citenamefont {Mac{\'e}}, \citenamefont {Vernier},\ and\ \citenamefont
  {Alet}}]{giraud_probing_2020}%
  \BibitemOpen
  \bibfield  {author} {\bibinfo {author} {\bibfnamefont {O.}~\bibnamefont
  {Giraud}}, \bibinfo {author} {\bibfnamefont {N.}~\bibnamefont {Mac{\'e}}},
  \bibinfo {author} {\bibfnamefont {E.}~\bibnamefont {Vernier}}, \ and\
  \bibinfo {author} {\bibfnamefont {F.}~\bibnamefont {Alet}},\ }\href
  {http://arxiv.org/abs/2008.11173} {\bibfield  {journal} {\bibinfo  {journal}
  {arXiv:2008.11173}\ } (\bibinfo {year} {2020})}\BibitemShut {NoStop}%
\bibitem [{Note3()}]{Note3}%
  \BibitemOpen
  \bibinfo {note} {We observe a critical power-law decay ${\protect \overline
  {|\delimiter "426830A \sigma _1^x\protect \tmspace +\thinmuskip
  {.1667em}\sigma _L^x\delimiter "526930B |}}\sim L^{-\omega _{\protect \rm
  b}}$ at $g_c= 0.24$ with $\omega _{\protect \rm b}\approx 0.17$ which
  contrasts with free-fermions where $\omega _b=1$~\cite
  {fisher_distributions_1998}.}\BibitemShut {Stop}%
\bibitem [{notexi()}]{notexi}%
  \BibitemOpen
  \bibinfo {note} {In the MBL PM regime, typical and average localization
  lengths show very similar behaviors.}\BibitemShut {Stop}%
\bibitem [{Note4()}]{Note4}%
  \BibitemOpen
  \bibinfo {note} {Where the typical localization length is $1/\delta $~\cite
  {fisher_critical_1995}.}\BibitemShut {Stop}%
\bibitem [{\citenamefont {Parameswaran}\ and\ \citenamefont
  {Vasseur}(2018)}]{Parameswaran_2018}%
  \BibitemOpen
  \bibfield  {author} {\bibinfo {author} {\bibfnamefont {S.~A.}\ \bibnamefont
  {Parameswaran}}\ and\ \bibinfo {author} {\bibfnamefont {R.}~\bibnamefont
  {Vasseur}},\ }\href {\doibase 10.1088/1361-6633/aac9ed} {\bibfield  {journal}
  {\bibinfo  {journal} {Reports on Progress in Physics}\ }\textbf {\bibinfo
  {volume} {81}},\ \bibinfo {pages} {082501} (\bibinfo {year}
  {2018})}\BibitemShut {NoStop}%
\bibitem [{\citenamefont {Rainis}\ and\ \citenamefont
  {Loss}(2012)}]{PhysRevB.85.174533}%
  \BibitemOpen
  \bibfield  {author} {\bibinfo {author} {\bibfnamefont {D.}~\bibnamefont
  {Rainis}}\ and\ \bibinfo {author} {\bibfnamefont {D.}~\bibnamefont {Loss}},\
  }\href {\doibase 10.1103/PhysRevB.85.174533} {\bibfield  {journal} {\bibinfo
  {journal} {Phys. Rev. B}\ }\textbf {\bibinfo {volume} {85}},\ \bibinfo
  {pages} {174533} (\bibinfo {year} {2012})}\BibitemShut {NoStop}%
\bibitem [{\citenamefont {Morningstar}\ \emph {et~al.}(2021)\citenamefont
  {Morningstar}, \citenamefont {Colmenarez}, \citenamefont {Khemani},
  \citenamefont {Luitz},\ and\ \citenamefont
  {Huse}}]{morningstar2021avalanches}%
  \BibitemOpen
  \bibfield  {author} {\bibinfo {author} {\bibfnamefont {A.}~\bibnamefont
  {Morningstar}}, \bibinfo {author} {\bibfnamefont {L.}~\bibnamefont
  {Colmenarez}}, \bibinfo {author} {\bibfnamefont {V.}~\bibnamefont {Khemani}},
  \bibinfo {author} {\bibfnamefont {D.~J.}\ \bibnamefont {Luitz}}, \ and\
  \bibinfo {author} {\bibfnamefont {D.~A.}\ \bibnamefont {Huse}},\ }\href
  {https://arxiv.org/abs/2107.05642} {} (\bibinfo {year} {2021}),\ \Eprint
  {http://arxiv.org/abs/2107.05642} {arXiv:2107.05642} \BibitemShut {NoStop}%
\bibitem [{\citenamefont {Sels}(2021)}]{sels2021markovian}%
  \BibitemOpen
  \bibfield  {author} {\bibinfo {author} {\bibfnamefont {D.}~\bibnamefont
  {Sels}},\ }\href {https://arxiv.org/abs/2108.10796} {} (\bibinfo {year}
  {2021}),\ \Eprint {http://arxiv.org/abs/2108.10796} {arXiv:2108.10796}
  \BibitemShut {NoStop}%
\bibitem [{\citenamefont {Fisher}\ and\ \citenamefont
  {Young}(1998)}]{fisher_distributions_1998}%
  \BibitemOpen
  \bibfield  {author} {\bibinfo {author} {\bibfnamefont {D.~S.}\ \bibnamefont
  {Fisher}}\ and\ \bibinfo {author} {\bibfnamefont {A.~P.}\ \bibnamefont
  {Young}},\ }\href {\doibase 10.1103/PhysRevB.58.9131} {\bibfield  {journal}
  {\bibinfo  {journal} {Phys. Rev. B}\ }\textbf {\bibinfo {volume} {58}},\
  \bibinfo {pages} {9131} (\bibinfo {year} {1998})}\BibitemShut {NoStop}%
\bibitem [{\citenamefont {Mi~{\it{et al.}}}(2021)}]{mi_observation_2021}%
  \BibitemOpen
  \bibfield  {author} {\bibinfo {author} {\bibfnamefont {X.}~\bibnamefont
  {Mi~{\it{et al.}}}},\ }\href {http://arxiv.org/abs/2107.13571} {\bibfield
  {journal} {\bibinfo  {journal} {arXiv:2107.13571}\ } (\bibinfo {year}
  {2021})}\BibitemShut {NoStop}%
\bibitem [{\citenamefont {Deutsch}(1991)}]{deutsch_quantum_1991}%
  \BibitemOpen
  \bibfield  {author} {\bibinfo {author} {\bibfnamefont {J.~M.}\ \bibnamefont
  {Deutsch}},\ }\href {\doibase 10.1103/PhysRevA.43.2046} {\bibfield  {journal}
  {\bibinfo  {journal} {Phys. Rev. A}\ }\textbf {\bibinfo {volume} {43}},\
  \bibinfo {pages} {2046} (\bibinfo {year} {1991})}\BibitemShut {NoStop}%
\bibitem [{\citenamefont {Srednicki}(1994)}]{srednicki_chaos_1994}%
  \BibitemOpen
  \bibfield  {author} {\bibinfo {author} {\bibfnamefont {M.}~\bibnamefont
  {Srednicki}},\ }\href {\doibase 10.1103/PhysRevE.50.888} {\bibfield
  {journal} {\bibinfo  {journal} {Phys. Rev. E}\ }\textbf {\bibinfo {volume}
  {50}},\ \bibinfo {pages} {888} (\bibinfo {year} {1994})}\BibitemShut
  {NoStop}%
\bibitem [{Note5()}]{Note5}%
  \BibitemOpen
  \bibinfo {note} {These oscillations can be understood in the limit $J_i\ll 1$
  where the $2^{\protect \rm nd} $neighbor terms $g\sigma _i^x\sigma _{i+2}^x$
  dominates over nearest-neighbor, thus strongly reducing odd-distant
  correlations.}\BibitemShut {Stop}%
\end{thebibliography}
%

\vskip 1cm
\begin{widetext}

\section{Supplemental material: spin correlations} Spontaneous breaking of $\mathbb Z_2$ symmetry associated to magnetic order with $\langle\sigma_i^x\rangle\neq 0$ only occurs in the thermodynamic limit. One can nevertheless probe it on finite chains using the squared spin correlator ${\cal{C}}_{i,j}=\langle \sigma_i^x \sigma_j^x\rangle^2$ (the square integrates out sign fluctuations at high-energy), preferring to avoid the spin-glass structure factor $\chi^{\rm SG}_{L}=\frac{1}{L}\sum_{i,j}{\cal{C}}_{i,j}$, more used~\cite{kjall_many-body_2014,vasseur_particle-hole_2016,kjall_many-body_2014,sahay_emergent_2021,mi_observation_2021}, but less precise. Indeed $\chi_{L}^{\rm SG}$  provides a reasonably good qualitative estimate for detecting SG order since it diverges with $L$, while it goes to a constant outside the quantum-ordered regime. However, there is no reason to expect a finite-size crossing of $\chi_{L}^{\rm SG}$ precisely at the critical point~\cite{vasseur_particle-hole_2016,kjall_many-body_2014,sahay_emergent_2021,mi_observation_2021} because the long distance decay of the critical correlations is not known. Moreover, the integrated form of $\chi_{L}^{\rm SG}$ enhances non-universal short-range effects, and it is safer to directly focus  on two-point correlators.

Typical pairwise correlators ${\cal{C}}^{\rm typ}(r)={\rm{e}}^{\overline{\ln\,{\cal{C}}_{i,i+r}}'}$ (the average ${\overline{[\cdots]}}'$ is taken over disordered samples and all available sites $i$) are shown in Fig.~\ref{fig:correlations}  for three representative situations. 
(a) Deep in the MBL SG regime, ${\cal{C}}^{\rm typ}(r)$ rapidly saturates to a roughly size-independent finite value, thus testifying for magnetic order. (b) In the middle of the thermal region we nicely verify the $2^{-L}$ behavior, expected from eigenstate thermalization hypothesis (ETH)~\cite{deutsch_quantum_1991,srednicki_chaos_1994}. 
(c) The MBL PM phase displays short-range correlations, exponentially decaying with additional oscillations~\footnote{These oscillations can be understood in the limit $J_i\ll 1$ where the $2^{\rm nd} $neighbor terms $g\sigma_i^x\sigma_{i+2}^x$ dominates over nearest-neighbor, thus strongly reducing odd-distant correlations.}.

\begin{figure}[ht!]
   \centering
   \includegraphics[width=.5\columnwidth,clip]{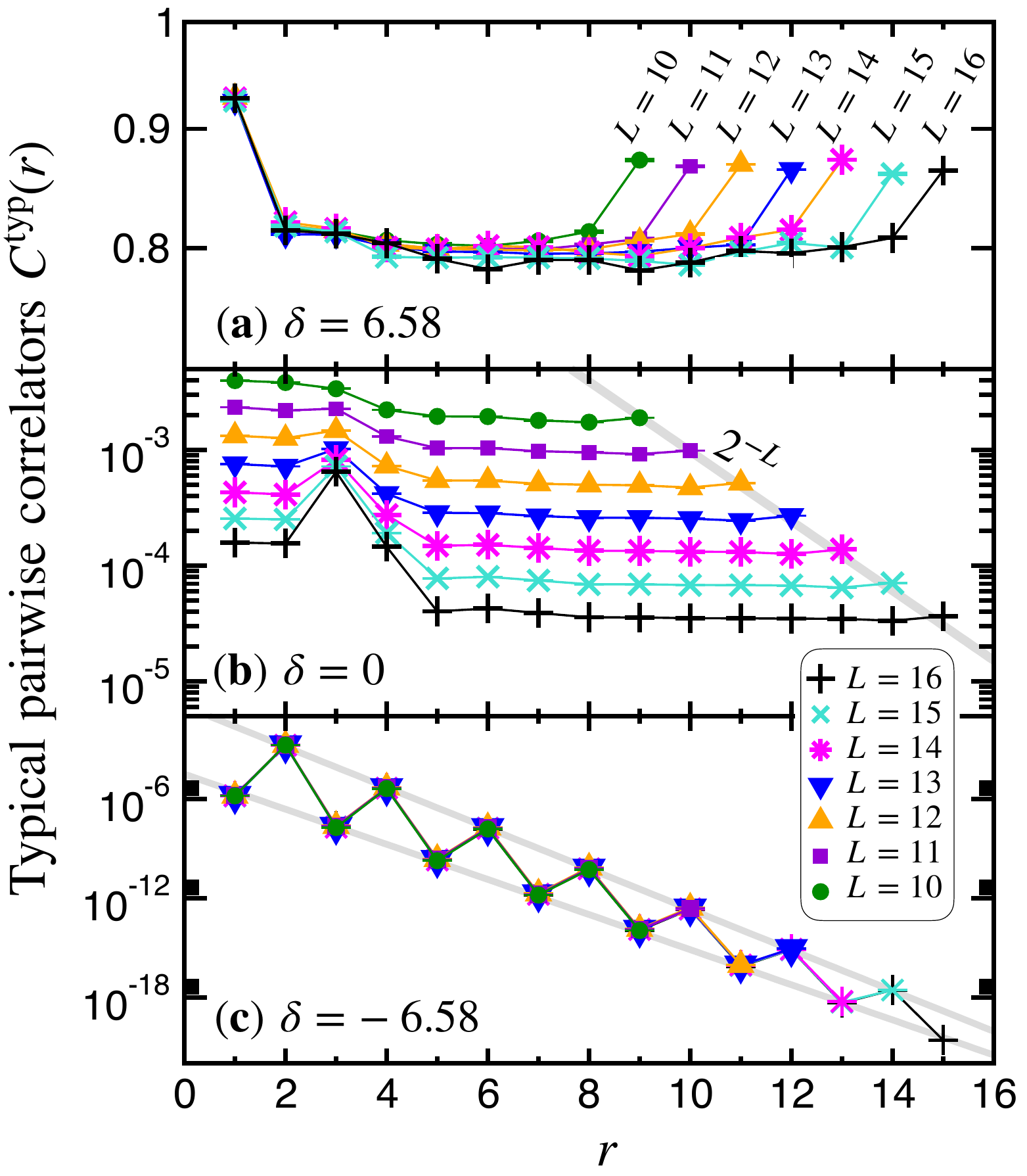}
      \caption{ED results for open IM chains at $g=0.5$, showing the decay of typical correlators ${\cal{C}}^{\rm typ}(r)$ for three representative cases. (a) In the MBL SG regime ($\delta=6.58$) spin-glass order is clear, with a notable boundary enhancement. (b) Ergodic ($\delta=0$): ETH behavior is evident after a few sites: ${\cal{C}}^{\rm typ}(r\gtrsim 5)\to 2^{-L}$ (grey line). (c) MBL PM ($\delta=-6.58$): we observe fast (even-odd oscillating  and size-independent) exponential decay ${\cal{C}}_0 \exp(-r/\xi)$, with fitting parameters $({\cal{C}}_0,\,\xi)_{\rm even}=(0.41,\,0.35)$ and  $({\cal{C}}_0,\,\xi)_{\rm odd}=(3.2\times 10^{-5},\, 0.408)$ for the available range (fits are shown by grey lines).}
   \label{fig:correlations}
\end{figure}
\end{widetext}
\end{document}